\documentclass[twoside,11pt]{article}
\usepackage{graphicx,subfigure}
\graphicspath{{./Images/}}
\usepackage{rotating}
\usepackage{url}%times,float,
\usepackage{amsmath,amsfonts,amssymb,amsthm} %,dsfont
\usepackage{tabularx,booktabs}
\usepackage{placeins}
\usepackage{graphicx}
\newcolumntype{H}{>{\setbox0=\hbox\bgroup}c<{\egroup}@{}}
\renewcommand{\v}[1]{\mathbf{#1}}
 
\newtheorem{theorem}{Theorem}

\newtheorem{definition}[theorem]{Definition}

\begin{document}
\title{Metric and non-metric proximity\\ transformations at linear costs}
\author{Andrej Gisbrecht\\
       Theoretical Computer Science\\
       University of Bielefeld\\
       Center of Excellence,\\
       Universit\"atsstrasse 21-23, \\
       33615, Bielefeld\\
       Frank-Michael~Schleif\footnote{Corresponding author: schleify@cs.bham.ac.uk}\\
       School of Computer Science\\	
       University of Birmingham\\
       Birmingham\\
       B15 2TT\\
       United Kingdom}
%\editor{}

\maketitle              % typeset the title of the contribution

\begin{abstract}%
Domain specific (dis-)similarity or proximity measures used e.g. in alignment algorithms of sequence data, are popular to analyze
         complex data objects and to cover domain specific data properties. \emph{Without an underlying vector space} these data are given as
	pairwise (dis-)similarities only. The few available methods for such data focus widely on similarities and do not scale to large data sets. 
	Kernel methods are very effective for \emph{metric similarity} matrices, also at large scale, but costly transformations are necessary starting 
	with non-metric (dis-) similarities. We propose an integrative combination of Nystr\"om approximation, potential double centering and 
         eigenvalue correction to obtain valid kernel matrices at \emph{linear costs} in the number of samples. By the proposed approach 
	effective kernel approaches, become accessible. Experiments with several larger (dis-)similarity data sets show that the proposed 
	method achieves much better runtime performance than the standard strategy while keeping competitive model accuracy. 
	The main contribution is an efficient and accurate technique, to convert (potentially non-metric) large scale \emph{dissimilarity matrices} 
	into approximated positive semi-definite kernel matrices at linear costs.
\end{abstract}


\section{Introduction}
In many application areas such as bioinformatics, text mining, image retrieval, spectroscopy domains or
social networks the available electronic data are increasing and get more complex in size and representation.
In general these data are not given in vectorial form and \emph{domain specific} (dis-)similarity measures 
are used, as a replacement or complement to Euclidean measures. These data are also often associated to dedicated 
structures which make a representation in terms of Euclidean vectors difficult: biological sequence data, text files, XML data, 
trees, graphs, or time series \cite{DBLP:journals/jmlr/ChenGGRC09,mediansom,neuhaus} are of this type. 
These data are inherently compositional and a feature representation leads to information loss. 
As an alternative, tailored dissimilarity measures such as pairwise  alignment functions, kernels for structures 
or other domain specific similarity and dissimilarity functions can  be used as the interface to the data.
But also for vectorial data,
non-metric proximity measures are common in some disciplines.
%blocking the direct way to effective kernel approaches.
An example of this type is the use of divergence measures \cite{Cichocki20101532}
which are very popular for spectral data analysis in chemistry, geo- and medical
sciences \cite{Schleif2010h,Nguyen2013691}, and are not metric in general. 
In such cases, machine learning techniques which can deal with pairwise 
non-metric similarities or dissimilarities are attractive \cite{Pekalska2005a}.

The paper is organized as follows. First we give a brief review of related work. Subsequently we review common
transformation techniques for dissimilarity data and discuss the influence of non-Euclidean measures, by eigenvalue corrections. 
Thereafter we discuss alternative methods for processing small dissimilarity data. We extend this discussion to approximation strategies and 
give an alternative derivation of the Nystr\"om approximation together
with a convergence proof, also for indefinite kernels.
This allows us to apply the Nystr\"om technique
to similarities as well as for dissimilarities.
Thus, we can link both strategies effectively to use kernel methods for the analysis of larger (non-)metric dissimilarity data.
Then we show the effectiveness of the proposed approach
by different supervised learning tasks aligned with various error measures. We also discuss differences and commons
to some known approaches supported by experiments on simulated data\footnote{This article contains extended and improved results and is based on \cite{Schleif2013b}}.

\section{Related work}
Similarity and dissimilarity learning or for short proximity learning has attracted wide attention over the last years, pioneered by work of \cite{Goldfarb1984575} and
major contributions in \cite{Pekalska2005a} and different other research groups. As will be detailed more formally in the next section,
the learning of proximities is challenging under different aspects: in general there is no underlying vector space, the proximities may be non-Euclidean,
the data may not be metric. As mentioned before a symmetric matrix of metric similarities between objects is essentially a kernel and can be analyzed
by a multitude of kernel methods \cite{Cristianini2004a}.  But complex preprocessing steps are necessary, as discussed in the following, 
to apply them on non-metric (dis-)similarities. Some recent work discussed non-metric \emph{similarities} in the context of kernel approaches
by means of indefinite kernels see e.g. \cite{Liwicki20121624,Haasdonk2009a}, resulting in non-convex formulations. Other approaches try to make the
kernel representation positive semi definite (psd) or learn an alternative psd proxy matrix close to the original one
\cite{DBLP:journals/jmlr/ChenGGRC09,DBLP:conf/icml/ChenGR09}, but with high computational costs. For dissimilarity matrices only few
approaches have been published \cite{Lu30082005,DBLP:journals/siammax/BrickellDST08} both with quadratic to cubic computational costs
in the number of samples. In fact, as discussed in the work of \cite{Pekalska2005a},
non-Euclidean proximities can encode important information in the Euclidean
as well as in non-Euclidean parts of space,
represented by the positive and negative eigenvalues
of the corresponding similarity matrix, respectively.
Thus, transformations of similarities to make them psd,
by e.g. truncating the negative eigenvalues,
may be inappropriate \cite{DBLP:conf/sspr/PekalskaDGB04}.
This however is very data dependent and for a large number of datasets negative eigenvalues may be actually noise effects
while for other data sets the negative eigenvalues carry relevant information \cite{DBLP:phd/de/Laub2004,DBLP:journals/pr/LaubRBM06}.
Often non-psd kernels are still used with kernel algorithms but actually on a heuristical basis, since corresponding error bounds are provided 
only for psd kernels in general. As we will see in the experiments for strongly non-psd data it may happen that standard kernel methods
fail to converge due to the violation of underlying assumptions.

Another strategy is to use a more general theory of learning with similarity functions proposed in \cite{DBLP:journals/ml/BalcanBS08}.
Which can be used to identify descriptive or discriminative models based on a available similarity function under some conditions
\cite{DBLP:conf/nips/KarJ12}. A practical approach of the last type for classification problems was provided in \cite{DBLP:conf/nips/KarJ11}. 
The model is defined on a fixed randomly chosen set of landmarks per class and a transfer function. Thereby the landmarks are a small set of columns
(or rows) of a kernel matrix which are used to formulate the decision function. The weights of the decision
function are then optimized by standard approaches. The results are however in general substantially worse than those provided in 
\cite{DBLP:journals/jmlr/ChenGGRC09} where the datasets are taken from.  
%The same authors extended this concept in \cite{DBLP:conf/nips/KarJ12<} to the problem of regression incorporating a sparse optimization strategy on i.i.d. sampled landmarks. 
%There the problem is formulated as a sparse linear regression problem. While very appealing the effectiveness of the approach for larger, 
%realistic data sets including outliers is not addressed and it is not clear if the proposed approach is superior to other sparse regression
%approaches using strategies of  \cite{DBLP:journals/jmlr/ChenGGRC09,DBLP:conf/fusion/ChenG09,DBLP:conf/icml/ChenGR09} to address non-psd similarities.

In the following we will focus on non-metric proximities and especially \emph{dissimilarities}. Native methods for the analysis of matrix dissimilarity data 
have been proposed in \cite{DBLP:journals/neco/GraepelO99,Pekalska2005a,DBLP:journals/jmlr/PekalskaPD01,Schleif2012k}, but are in general based on non-convex optimization schemes 
and with quadratic to linear memory and runtime complexity, the later employing some of the approximation techniques discussed subsequently 
and additional heuristics. The strategy to correct non-metric dissimilarities is addressed in the literature only for smaller data sets.  And there exist
basically three approaches to make them metric. The first one is to modify the (symmetric) dissimilarity matrix such that all triangle equations
in the data are fulfilled \cite{DBLP:journals/siammax/BrickellDST08}, which is called the \emph{metric-nearness} problem. The second strategy
is to learn again a metric proxy matrix \cite{Lu30082005}. Both strategies are quite costly and not used at large scale. The third approach 
is based on converting the dissimilarities to similarities, by double centering followed by an eigenvalue correction of the similarities and
back conversion to dissimilarities. These steps scale quadratic and cubic, respectively. We focus on the last approach and provide a runtime
and memory efficient solution for problems at large scale\footnote{With large we refer to a sample size $N \in [1e3-1e6]$. We do not focus
on very big data - which are (not yet) considered in the area of proximity learning.}. 

The approximation concepts used in the following are based on the Nystr\"om approximation which was introduced to machine learning by the work of
\cite{DBLP:conf/nips/WilliamsS00}. In \cite{DBLP:journals/pami/FowlkesBCM04} the Nystr\"om approximation was used to simplify the normalized Cut
problem, which can be considered as a clustering problem. This work was however valid for \emph{psd similarity} matrices, only. An extension to
\emph{non-psd} similarities was addressed in \cite{DBLP:conf/eccv/BelongieFCM02}, but the derivation can still lead to an invalid matrix approximation
\footnote{The derivation of $Z$ on p 535 for negative eigenvalues in $\Lambda$ leads to complex values and hence invalid results.
However the strategy proposed in the corresponding paper may have removed the negative eigenvalues in $\Lambda$, due to a rank reduction,
explaining the experimental results. But the cut-off of negative eigenvalues can again be criticized \cite{DBLP:conf/sspr/PekalskaDGB04}}. 
Our proposal derives valid eigenvector and eigenvalue estimates also for non-psd proximity matrices.

Large (dis-)similarity data are common in biology like the famous
\emph{UniProt-/SwissProt}-database with $\approx 500,000$ entries
or \emph{GenBank} with $\approx 135,000$ entries, but there are many more (dis-)similarity data as discussed in the work based 
on \cite{Pekalska2005a,DBLP:journals/tsmc/PekalskaD08}. These growing data sets request effective and generic modeling approaches.

Here we will show how potentially non-metric (dis-)simi\-larities can be effectively processed by standard kernel methods by
correcting the proximity data with linear costs.
%which, to the authors best knowledge has not been reported before. 
The proposed strategies permit the effective application of many kernel methods for these type of data under very mild conditions. 

Especially for metric dissimilarities the approach keeps the known guarantees, like generalization bounds (see e.g. \cite{DBLP:journals/jmlr/DrineasM05}).
For non-psd data we give a convergence proof, but the corresponding bounds are still open, yet our experiments are promising.

\section{Transformation techniques for (dis-)similarities}
\label{sec:trafos}
Let $\v{v}_j \in  \mathbb{V}$ be a set of objects defined in some data space, with $|\mathbb{V}|=N$. 
We assume, there exists a dissimilarity measure such that $\mathbf{D} \in \mathbb{R}^{N \times N}$ is a dissimilarity matrix 
measuring the pairwise dissimilarities $D_{ij}=d(\v{v}_j,\v{v}_i)^2$
between all pairs $(\v{v}_i,\v{v}_j) \in \mathbb{V}$
\footnote{We assume $D_{ij}$ to be squared to simplify the notation.}.
Any reasonable (possibly non-metric) distance measure  is sufficient. 
We assume zero diagonal $d(\v{v}_i,\v{v}_i)=0$ for all $i$ and symmetry $d(\v{v}_i,\v{v}_j)=d(\v{v}_j,\v{v}_i)$
for all $i,j$.

\subsection{Transformation of dissimilarities and similarities into each other}
Every dissimilarity matrix $\mathbf{D}$ can be seen as a distance matrix
computed in some, not necessarily Euclidean, vector space.
The matrix of the inner products computed in this space
is the corresponding similarity matrix $\mathbf{S}$.
It can be computed from $\mathbf{D}$ directly
by a process referred to as double centering \cite{Pekalska2005a}:
\begin{eqnarray*}
	\mathbf{S} &=& -\mathbf{J} \mathbf{D} \mathbf{J}/2 \\
	\mathbf{J} &=& (\mathbf{I}-\mathbf{1}\mathbf{1}^\top/N)
\end{eqnarray*}
with identity matrix $\mathbf{I}$  and vector of ones $\mathbf{1}$.
Similarly, it is possible to construct the dissimilarity matrix element-wise
from the matrix of inner products $\mathbf{S}$
\[D_{ij} = S_{ii} + S_{jj} - 2 S_{ij}.\]
As we can see, both matrices $\mathbf{D}$ and $\mathbf{S}$
are closely related to each other and represent the same data,
up to translation, which is lost by the double-centering step. 
If the mean estimate, used in the double centering step, is inaccurate
the conversion of $\mathbf{D}$ to $\mathbf{S}$ is inaccurate as well, which
can have a negative impact on e.g. a classifier based on $\mathbf{S}$.

The data stems from an Euclidean space,
and therefore the distances $d_{ij}$ are Euclidean,
if and only if $\mathbf{S}$ is positive semi-definite (psd) \cite{Berg1984}.
This is the case, when we observe only non-negative eigenvalues
in the eigenspectrum of the matrix $\mathbf{S}$ associated to $\mathbf{D}$.
Such psd matrices $\mathbf{S}$ are also referred to as kernels
and there are many classification techniques,
which have been proposed to deal with such data,
like the support vector machine (SVM) \cite{vapnik2000nature}. 
In the case of non-psd similarities, the mercer kernel based techniques
are no longer guaranteed to work properly
and additional transformations of the data are required
or the methods have to be modified substantially, 
effecting the overall runtime efficiency or desired properties like convexity \cite{Ong2004639,Haasdonk2005482}.
To define these transformations we need first
to understand the pseudo-Euclidean space.

%or indefinite kernel matrices,
%preprocessings as described in the following are \emph{required to guarantee} psd.
\subsection{Pseudo-Euclidean embedding}\label{sec:pseudo_eucl_embedd}
Given a symmetric dissimilarity with zero diagonal,
an embedding of the data in a pseudo-Euclidean vector space
%determined by the eigenvector decomposition of the associated matrix $\mathbf{S}$
is always possible \cite{Goldfarb1984575}.

\begin{definition}[Pseudo-Euclidean space \cite{Pekalska2005a}]
A pseudo-Euclidean space $\xi=\mathbb{R}^{(p,q)}$
is a real vector space equipped with a non-degenerate,
indefinite inner product $\langle .,.\rangle_\xi$. $\xi$ admits a direct orthogonal decomposition $\xi=\xi_+  \oplus \xi_-$ where
$\xi_+= \mathbb{R}^p$ and $\xi_-= \mathbb{R}^q$ and the inner product is positive definite on $\xi_+$ and negative
definite on $\xi_-$. The space $\xi$ is therefore characterized by the signature $(p,q)$.
\end{definition}

A symmetric bi-linear form in this space is given by 
\[
\langle\v{x},\v{y}\rangle_{p,q} =
\sum_{i=1}^p x_i y_i - \sum_{i=p+1}^{p+q} x_i y_i =
\v{x}^\top \mathbf{I}_{p,q}\v{y}
\]
where $\mathbf{I}_{p,q}$ is a diagonal matrix with $p$ entries $1$ and $q$ entries $-1$.
Given the eigendecomposition of a similarity matrix
$\mathbf{S} = \mathbf{U} \mathbf{\Lambda} \mathbf{U}^\top$
we can compute the corresponding vectorial representation $\mathbf{V}$
in the pseudo-Euclidean space by
\begin{equation}
\mathbf{V} = \mathbf{U}_{p+q} \left|\mathbf{\Lambda}_{p+q}\right|^{1/2}
\label{eq:embedding}
\end{equation}
where $\mathbf{\Lambda}_{p+q}$ consists of $p$ positive and $q$ negative
non-zero eigenvalues and $\mathbf{U}_{p+q}$ consists of the corresponding eigenvectors.
It is straightforward to see that
$D_{ij}=\langle\v{v}_i-\v{v}_j,\v{v}_i-\v{v}_j\rangle_{p,q}$
holds for every pair of data points.
Similarly to the signature $(p, q)$ of a space $\xi$,
we describe our finite data sets, given by a matrix $\mathbf{D}$ or $\mathbf{S}$,
by the extended signature $(p, q, N-p-q)$
which represents the number of positive eigenvalues $p$,
the number of negative eigenvalues $q$
and the number of the remaining zero eigenvalues
in the similarity matrix.

\subsection{Dealing with pseudo-Euclidean data}
\label{sec:trafos_corr}
In \cite{DBLP:journals/jmlr/ChenGGRC09} different strategies were analyzed to
obtain valid kernel matrices for a given similarity matrix $\mathbf{S}$,
most popular are: \emph{flipping, clipping, vector-representation, shift correction}.
The underlying idea is to remove negative eigenvalues
in the eigenspectrum of the matrix $\mathbf{S}$.
One may also try to learn an alternative psd kernel representation with maximum alignment to the original non-psd kernel matrix \cite{DBLP:journals/jmlr/ChenGGRC09,DBLP:conf/icml/ChenGR09,DBLP:journals/jmlr/LiZY09}
or split the proximities based on positive and negative eigenvalues as discussed in
\cite{Pekalska2005a,Haasdonk2009a}.

The \emph{flip}-operation takes the absolute eigenvalues of the matrix $\mathbf{S}$.
This corresponds to ignoring the separation of the space $\xi$
into $\xi_+$ and $\xi_-$ and instead computing in the space $\mathbb{R}^{p+q}$.
This approach preserves the variation in the data
and could be revoked for some techniques after the training
by simply reintroducing the matrix $\mathbf{I}_{p,q}$ into the inner product.

The \emph{shift}-operation increases all eigenvalues by the absolute
value of the minimal eigenvalue.
This approach performs a non-linear transformation in the pseudo-Euclidean space,
emphasizing $\xi_+$ and nearly eliminating $\xi_-$.

The \emph{clip}-operation sets all negative eigenvalues to zero.
This approach corresponds to ignoring the space $\xi_-$ completely.
As discussed in \cite{DBLP:conf/sspr/PekalskaDGB04}, depending on the data set,
this space could carry important information
and removing it would make some tasks, as e.g. classification, impossible.

After the transformation of the eigenvalues,
the corrected matrix $\mathbf{S}^*$ is obtained as
$\mathbf{S}^* = \mathbf{U} \mathbf{\Lambda}^* \mathbf{U}^\top$,
with $\mathbf{\Lambda}^*$ as
the modified eigenvalue matrix using one of the above operations.
The obtained matrix $\mathbf{S}^*$ can now be considered
as a valid kernel matrix $\mathbf{K}$
and kernel based approaches can be used to operate on the data.

The analysis in \cite{DBLP:conf/sspr/PekalskaDGB04} indicates that for non-Euclidean dissimilarities some corrections like above 
may change the data representation such that information loss occurs.
This however is not yet systematically explored and very data dependent,
best supported by domain knowledge about the data or the used proximity measure.

Alternatively, techniques have been introduced which directly deal with possibly non-metric dissimilarities.
Using the Equation \eqref{eq:embedding} the data can be embedded into
the pseudo-Euclidean space.
Classical vectorial machine learning algorithms can then be adapted
to operate directly in the pseudo-Euclidean space.
This can be achieved by e.g. defining a positive definite
inner product in the space $\xi$.
Variations of this approach are also possible
whereby an explicit embedding is not necessary
and the training can be done implicitly,
based on the dissimilarity matrix only \cite{Pekalska2005a}.
A further strategy is to employ so called relational or proximity learning methods as discussed e.g. in \cite{Schleif2012k}.
The underlying models consist of prototypes,
which are implicitly defined as a weighted linear combination of training points: 
\begin{equation*}
	\v{w}_j=\sum_i\alpha_{ji}\v{v}_i\mbox{ with } \sum_i\alpha_{ji}=1\,. \qquad \mathbb{W}=\{\v{w}_1, \hdots, \v{w}_c\}
\end{equation*}
But this explicit representation is not necessary because the algorithms are based only 
on a specific form of distance calculations using the matrix $\mathbf{D}$ and
the potentially unknown vector space $V$ is not needed.
The basic idea is an implicit computation of distances $d(\cdot,\cdot)$
during the model calculation based on the dissimilarity matrix $\mathbf{D}$ using weights $\alpha$:
\begin{equation}
	d(\v{v}_i,\v{w}_j)^2=
	[\mathbf{D}\cdot\alpha_j]_i-\frac{1}{2}\cdot\alpha_j^\top \mathbf{D}\alpha_j
	\label{eq:rel_distance}.
\end{equation}
details in \cite{Schleif2012k}.
As shown e.g. in \cite{DBLP:journals/neco/HammerH10} the mentioned methods 
do not rely on a metric dissimilarity matrix $\mathbf{D}$,
but it is sufficient to have a symmetric $\mathbf{D}$ in a pseudo-Euclidean space,
with constant self-dissimilarities.

The \emph{dissimilarity space} approach is another technique
which does not embed the data into the pseudo-Euclidean space \cite{Pekalska2005a}.
Instead, one selects a representative set of points $\v{w}_i \in \mathbb{W}$
and considers for every point the dissimilarities to the set $\mathbb{W}$
as features, resulting in a vectorial representation
$\mathbf{x}_i=[d(\v{v}_i,\v{w}_1),d(\v{v}_i,\v{w}_2),d(\v{v}_i,\v{w}_3),...]^\top$.
This corresponds to an embedding into an Euclidean space
with the dimensionality equal to the size of the selected set of points.
These vectors can then be processed using any vectorial approaches.
A negative point of this representation is the
change of the original data representation
which may disturb the structure of the data.
It is also highly reliable on a good representative set,
since highly correlated sampled points generate similar features
and the correlation information is lost in the embedded space.

\subsection{Complexity}
The methods discussed before are suitable for data analysis based on similarity or dissimilarity data
where the number of samples $N$ is rather small, e.g. scales by some thousand samples. 
For large $N$, most of the techniques discussed above become infeasible.
All techniques which use the full (dis-)similarity matrix,
have $\mathcal{O}(N^2)$ memory complexity
and thus at least $\mathcal{O}(N^2)$ computational complexity.

Double centering, if done naively, is cubic,
although after simplifications it can be computed in $\mathcal{O}(N^2)$.
Transformation from $\mathbf{S}$ to $\mathbf{D}$ can be done element-wise,
but if the full matrix is required it is still quadratic.

All the techniques relying on the full eigenvalue decomposition,
e.g. for eigenvalue correction or for explicit pseudo-Euclidean embedding,
have an $\mathcal{O}(N^3)$ computational complexity.

The only exception is the dissimilarity space approach.
If it possible to select a good representative set of a small size,
one can achieve linear computational and memory complexity.
The technique becomes quadratic as well,
if all data points are selected as the representative set.

Other then this,
only for \emph{metric, similarity data} (psd kernels) efficient approaches have been proposed before, e.g. 
the Core-Vector Machine (CVM) \cite{DBLP:conf/icml/TsangKK07} or low-rank linearized SVM \cite{DBLP:journals/jmlr/ZhangLWM12}
for classification problems or an approximated kernel k-means algorithm for clustering \cite{DBLP:conf/kdd/ChittaJHJ11}.

\begin{figure}
	\centering
	\includegraphics[width=0.8\columnwidth]{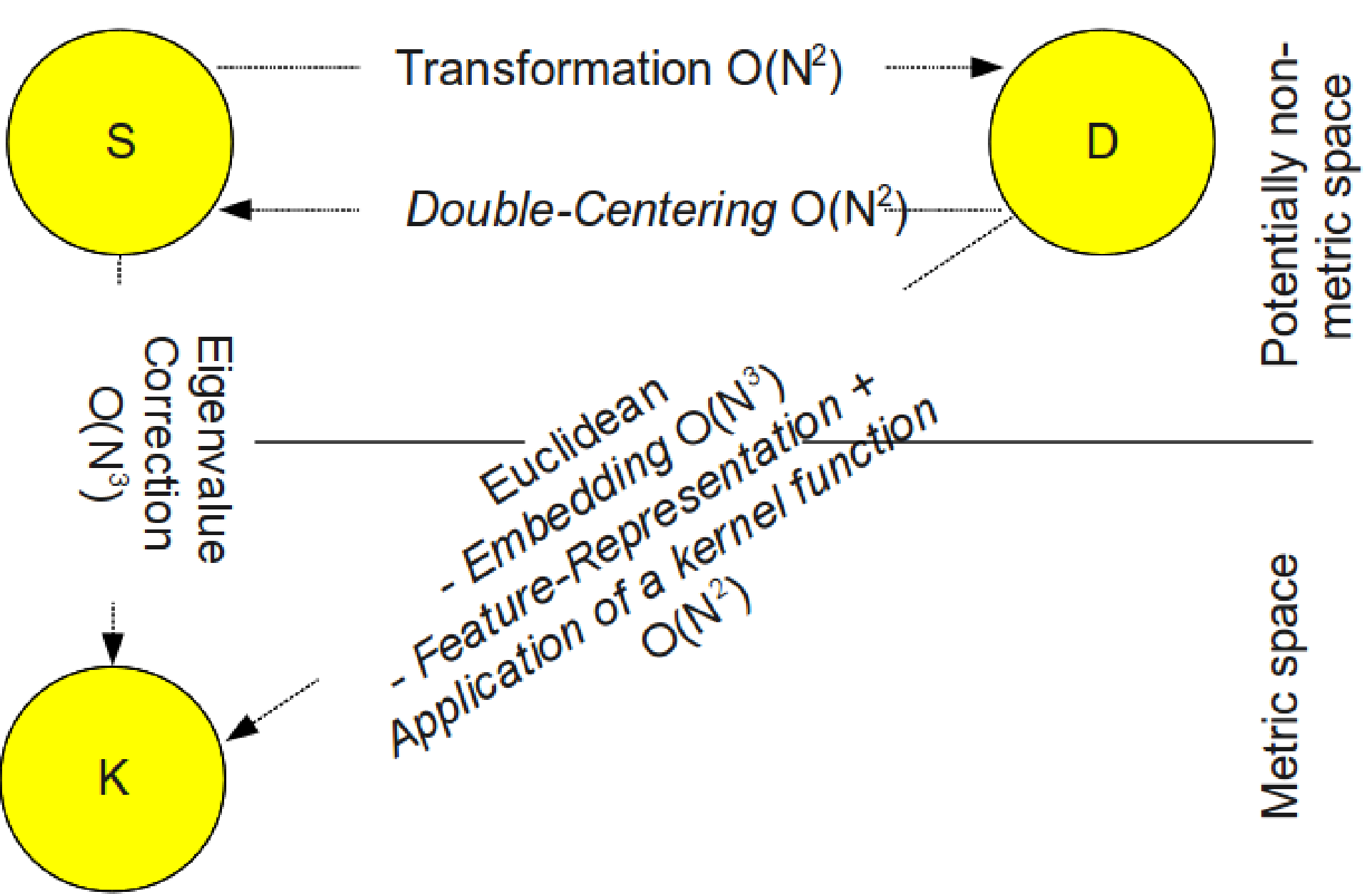}
	\caption{Schema of the relation between similarities and dissimilarities.}
	\label{fig:simdis_schema}
\end{figure}

A schematic view of the relations between $\mathbf{S}$ and $\mathbf{D}$ and its transformations
%\footnote{Transformation equations are given also in the following sections.} 
is shown in Figure \ref{fig:simdis_schema}, including the complexity of the transformations.
Some of the steps can be done more efficiently by known methods,
but with additional constraints or in atypical settings as discussed in the following.

In the following, we discuss techniques to deal with larger sample sets
for potentially non-metric similarity and especially dissimilarity data.
We show how standard kernel methods can be used,
assuming that for non-metric data,
the necessary transformations have no severe negative
influence on the data accuracy. Basically also core-set techniques \cite{DBLP:journals/comgeo/BadoiuC08}
become accessible for large potentially non-metric (dis-)similarity data in this way, but at the cost of multiple additional intermediate steps.
In particular, we investigate the Nystr\"om approximation technique,
as low rank linear time approximation technique;
we will show its suitability and linear time complexity
for similarities as well as dissimilarities,
applied on the raw data as well as for the eigenvalue correction.

\section{Nystr\"om approximation}
\label{sec:ny_approx}
As shown in \cite{DBLP:conf/nips/WilliamsS00}, given a symmetric positive semi-definite kernel matrix $\mathbf{K}$,
it is possible to create a low rank approximation of this matrix
using the Nystr\"om technique \cite{ny_orig}.
The idea is to sample $m$ points, the so called landmarks,
and to analyze the small $m \times m$ kernel matrix $\mathbf{K}_{m,m}$
constructed from the landmarks.
The eigenvalues and eigenvectors from the matrix $\mathbf{K}_{m,m}$
can be used to approximate the eigenvalues and eigenvectors
of the original matrix $\mathbf{K}$.
This allows to represent the complete matrix in terms
of a linear part of the full matrix only.
The final approximation takes the simple form
\begin{equation}
\mathbf{\hat{K}}=\mathbf{K}_{N,m} \mathbf{K}^{-1}_{m,m} \mathbf{K}_{m,N},
\label{eq:Ny_equation}
\end{equation}
where $\mathbf{K}_{N,m}$ is the kernel matrix between $N$ data points and $m$ landmarks
and $\mathbf{K}^{-1}_{m,m}$ is the Moore-Penrose pseudoinverse of the small matrix.

This technique has been proposed
in the context of Mercer kernel methods in \cite{DBLP:conf/nips/WilliamsS00}
with related proofs and bounds given in \cite{DBLP:journals/jmlr/DrineasM05}
and very recent results in \cite{DBLP:journals/corr/abs-1303-1849}.
It can be applied in conjunction with algorithms using the kernel matrix
in multiplications with other matrices or vectors only.
Due to the explicit low rank form as in Equation \eqref{eq:Ny_equation}
it is possible to select the order of multiplication,
thus reducing the complexity
from quadratic in the number of data points to a linear one.

\subsection{Eigenvalue decomposition of a Nystr\"om approximated matrix}\label{sec:eval_decomp}
In some applications it might be useful to compute the exact eigenvalue decomposition
of the approximated matrix $\mathbf{\hat{K}}$,
e.g. to compute the pseudo-inverse of this matrix.
We will show now, how this decomposition can be computed in linear time
\footnote{A similar strategy was used to construct large eigenmaps from \emph{psd} similarity matrices 
as recently shown \cite{JMLR:v14:talwalkar13a} but our approach applies also to non-psd matrices.}.
The psd matrix approximated by Equation \eqref{eq:Ny_equation} 
can be written as
\begin{align*}
\mathbf{\hat{K}} & = \mathbf{K}_{N,m} \mathbf{K}^{-1}_{m,m} \mathbf{K}_{m,N}\\
& = \mathbf{K}_{N,m} \mathbf{U} \mathbf{\Lambda}^{-1}
  \mathbf{U}^\top \mathbf{K}_{N,m}^\top\\
& = \mathbf{B} \mathbf{B}^\top,
\end{align*}
where we defined $\mathbf{B}=\mathbf{K}_{N,m} \mathbf{U} \mathbf{\Lambda}^{-1/2}$
with $\mathbf{U}$ and $\mathbf{\Lambda}$ being the eigenvectors and eigenvalues
of $\mathbf{K}_{m,m}$, respectively.
Further it follows
\begin{align*}
\mathbf{\hat{K}}^2 & = \mathbf{B} \mathbf{B}^\top \mathbf{B} \mathbf{B}^\top\\
& = \mathbf{B} \mathbf{V} \mathbf{A} \mathbf{V}^\top \mathbf{B}^\top,
\end{align*}
where $\mathbf{V}$ are the orthonormal eigenvectors of the matrix 
$\mathbf{B}^\top \mathbf{B}$
and $\mathbf{A}$ the matrix of its eigenvalues.
The corresponding eigenequation can be written as
$
\mathbf{B}^\top \mathbf{B} \mathbf{v} = a \mathbf{v}.
$
Multiplying it with $\mathbf{B}$ from left
we get the eigenequation for $\mathbf{\hat{K}}$
\[
\mathbf{B} \mathbf{B}^\top (\mathbf{B} \mathbf{v})
= a \left( \mathbf{B} \mathbf{v} \right).
\]
It is clear, that $\mathbf{A}$ must be the matrix of eigenvalues of $\mathbf{\hat{K}}$.
The matrix $\mathbf{B} \mathbf{v}$ is the matrix of the corresponding eigenvectors,
which are orthogonal but not necessary orthonormal.
The normalization can be computed from the decomposition
\begin{align*}
\mathbf{\hat{K}} & = \mathbf{B} \mathbf{V} \mathbf{V}^\top \mathbf{B}^\top\\
& = \mathbf{B} \mathbf{V} \mathbf{A}^{-1/2} \mathbf{A}
  \mathbf{A}^{-1/2} \mathbf{V}^\top \mathbf{B}^\top\\
& = \mathbf{C} \mathbf{A} \mathbf{C}^\top,
\end{align*}
where we defined $\mathbf{C} = \mathbf{B} \mathbf{V} \mathbf{A}^{-1/2}$
as the matrix of orthonormal eigenvectors of $\mathbf{\hat{K}}$.
Thus,
$
\mathbf{\hat{K}} = \mathbf{C} \mathbf{A} \mathbf{C}^\top
$
is the orthonormal eigendecomposition of $\mathbf{\hat{K}}$.

\subsection{Convergence proof}
The Nystr\"om approximation was proposed for the psd matrices
and thus, it was not accessible for distance matrices
and similarities coming from non-psd kernel functions.
First developments to apply the Nystr\"om technique to indefinite matrices
were presented in \cite{nips10gismokham,Schleif2012k}.
Although supported with experiments, a formal proof was lacking.
Here we present a proof, that shows, that the Nystr\"om approximated, 
possible indefinite, kernel converges in the operator norm to the true underlying kernel
as long as the number of landmarks is large enough. Generalization bounds will be a subject of future work.

Let $K$ be an integral operator and its kernel
$k\in L^2(\Omega^2)$ be a continuous symmetric function (not necessarily psd, i.e. it does not have to reproduce a Hilbert space):
\[
K f(x) := \int_\Omega k(x,y)f(y) d\mu(y).
\]
Without loss of generality let $\Omega$ be an interval $[a,b]\subset \mathbb{R}$
with measure 1.
Then $K$ is a compact operator in a Hilbert space $\mathfrak{H}$
\[
\|K\|_{L^2 \to L^2} :=
  \sup_{\|f\|\leq 1} \|K f\|_{L^2}
\leq \|k\|_{L_2},
\]
with the operator norm $\|.\|_{L^2 \to L^2}$ and the $L_2$-norm $\|.\|_{L_2}$.

We define a measurement operator $T_m$
which divides the space $\Omega$ into $m$ spaces $\Omega_j$,
each with the measure $1/m$.
% $\mu(\Omega_j)$.
%\frac{1}{\mu(\Omega_j)} 
It converts functions $f \in \mathfrak{H}$
to functions $f_m \in \mathfrak{H}_m$ which are piece-wise constant on each $\Omega_j$.
The corresponding integral kernel of $T_m$ is defined as:
\[
t_m(x,y):=
\begin{cases}
m & x, y \in \Omega_j \text{ for any } j \\
0 & \text{else}.
\end{cases}
\]
It follows for an $x \in \Omega_j$ that
\[
T_m f(x) = \int_\Omega t_m(x,y) f(y) d\mu(y) = m
\int_{\Omega_j} f(y) d\mu(y),
\]
where we can see, that the right hand side is 
the mean value of $f(y)$ on $\Omega_j$
and thus constant for all $x \in \Omega_j$.
%independent of $x$ and thus constant on $\Omega_j$.
This way, the operator $T_m$ allows us to approximate a function $f(x)$
by measuring it at $m$ places $f(x_j)$ and assuming that it is constant in between.
Measuring the operator $K$ we get $K_m := T_m \circ K$ with the integral kernel
\begin{align*}
\int_\Omega t_m(x,z) k(z,y) d\mu(z)
& = \sum_{j=1}^m \int_{\Omega_j} t_m(x,z) k(z,y) d\mu(z) \\
& = \sum_{j=1}^m 1_{\Omega_j}(x) m \int_{\Omega_j} k(z,y) d\mu(z) \\
& = \sum_{j=1}^m 1_{\Omega_j}(x) k_j(y) \\
& =: k_m(x,y),
\end{align*}
where $1_{\Omega_j}(x)$ is the indicator function which is $1$ if $x \in \Omega_j$
and $0$ elsewhere and we defined
$k_j=m \int_{\Omega_j} k(z,y) d\mu(z)$.

\noindent We can now analyze the convergence behavior of $K_m$ to $K$.
$\forall x \in \Omega_j$ and $\forall y \in \Omega$ we get
\begin{align*}
& \left| k_m(x,y) - k(x,y) \right| =\\
& = \left| m \int_{\Omega_j} k(z,y) d\mu(z)
 - m \int_{\Omega_j} k(x,y) d\mu(z) \right| \\
& \leq m \int_{\Omega_j} \left| k(z,y) - k(x,y) \right| \, d\mu(z).
\end{align*}
Since $k$ is continuous on the interval $[a,b]$, it is uniformly continuous
and we can bound
\begin{align*}
\left| k(z,y) - k(x,y) \right|
& \leq \mathcal{D}(\Omega_j)
 := \sup_{\substack{x_1, \, x_2 \in \Omega_j\\ y \in \Omega}}
  \left| k(x_1,y) - k(x_2,y) \right| \\
& \leq \delta_m := \max_j \mathcal{D}(\Omega_j)
\end{align*}
and therefore
\[
\sup_{\substack{x \in \Omega \\ y \in \Omega}}\left| k_m(x,y) - k(x,y) \right|
\leq \delta_m.
\]
For $m \to \infty$ the $\Omega_j$ become smaller and $\delta_m \to 0$,
thus kernel $k_m$ converges to $k$.
For the operators $K$ and $K_m$ it follows
\[
\| K_m - K \|_{L^2 \to L^2} \to 0
\]
which shows that $K_m$ converges to $K$ in the operator norm,
if the number of measurements goes to infinity.

\noindent Applying $K_m$ on $f$ results in
\begin{align*}
K_m f(x)
& = \int_{\Omega} k_m(x,y) f(y) d\mu(y) \\
& = \sum_{j=1}^m 1_{\Omega_j}(x) \int_{\Omega} k_j(y) f(y) d\mu(y) \\
& = \sum_{j=1}^m a_j 1_{\Omega_j}(x)
\end{align*}
where $a_j:=\int_{\Omega} k_j(y) f(y) d\mu(y)$ is a constant with respect to $x$.
It is clear that $K_m f$ is always in the linear hull of
$1_{\Omega_1}(x),...,1_{\Omega_m}(x)$ and the image of the operator
$\Im K_m=\operatorname{span}\{1_{\Omega_1}(x),...,1_{\Omega_m}(x)\}$ is $m$ dimensional.
Since the coefficients $a_j$ are finite, $K_m$ is a compact operator
and because the sequence of $K_m$ converges to $K$,
we see that $K$ is in fact a compact operator.

According to the "Perturbation of bounded operators" theorem \cite{DBLP:conf/colt/LuxburgBB04},
if a sequence $K_m$ converges to $K$ in the operator norm,
then for an isolated eigenvalue $\lambda$ of $K$
there exist isolated eigenvalues $\lambda_m$ of $K_m$
such that $\lambda_m \to \lambda$
and the corresponding spectral projections converge in operator norm.
%The same convergence applies to the corresponding eigenvectors.
This theorem allows us to estimate the eigenvalues and eigenfunctions of
the unknown operator $K$ by computing the eigendecomposition
of the measured operator $K_m$.

The eigenfunctions and eigenvalues of the operator $K_m$
are given as the solutions of the eigenequation
\begin{equation}
K_m f=\lambda f.
\label{eig_eq}
\end{equation}
We know that the left hand side of the equation is in the image of $K_m$
and therefore an eigenfunction $f$ must have the form
\begin{equation}
f(x)=\sum_{i=1}^m f_i 1_{\Omega_i}(x)
\label{eig_func}
\end{equation}
where $f_i$ are constants.
For the left side of the Equation \eqref{eig_eq} it follows
\begin{align*}
K_m f(x)
& = \int_{\Omega} \sum_{j=1}^m 1_{\Omega_j}(x) k_j(y) f(y) d\mu(y) \\
& = \sum_{j=1}^m 1_{\Omega_j}(x) 
\int_{\Omega} k_j(y) \sum_{i=1}^m f_i 1_{\Omega_i}(y) d\mu(y) \\
& = \sum_{j=1}^m \sum_{i=1}^m 1_{\Omega_j}(x) f_i 
\int_{\Omega_i} k_j(y) d\mu(y) \\
& = \sum_{j=1}^m \sum_{i=1}^m 1_{\Omega_j}(x) \frac{1}{m}
f_i k_{ji}
\end{align*}
and we defined
$k_{ji}=m \int_{\Omega_i} k_j(y) d\mu(y)
=m^2 \int_{\Omega_i} \int_{\Omega_j} k(y,z) d\mu(y) d\mu(z)$
which represents our measurement of the kernel $k$ around the $i$-th and $j$-th points.
If we combine the above equation with the Equation \eqref{eig_eq} for an $x \in \Omega_j$
we get
\[
\sum_{i=1}^m \frac{1}{m} k_{ji} f_i = \lambda f_j.
\label{eig_eq_j}
\]
This equation is a weighted eigenequation and we can turn it into a regular eigenequation
%by assuming $\mu(\Omega_i)=1/m \; \forall i$ and
%by defining $\tilde{k}_{ij}=\sqrt{\mu(\Omega_i)\mu(\Omega_j)} k_{ij}$
%and $\tilde{f}_i = \sqrt{\mu(\Omega_i)} f_i$.
by defining $\tilde{\lambda}=m\lambda$
and $\tilde{f}_i = f_i/\sqrt{m}$.
Thus, we get
\[
\sum_{i=1}^m k_{ji} \tilde{f}_i = \tilde{\lambda} \tilde{f}_j.
\]
Hence $\tilde{\lambda}$ and $\tilde{f}$ are the eigenvalues and eigenvectors
of matrix $(k_{ji})$.
Note, that $f_i$ are scaled to guarantee the normalization of $\tilde{f}$
\begin{align*}
1
& = \int_{\Omega} f(x) f(x) d\mu(x) \\
& = \int_{\Omega} \sum_{i=1}^m f_i^2 1_{\Omega_i}(x) d\mu(x)\\
& = \sum_{i=1}^m f_i^2 \int_{\Omega_i} d\mu(x)\\
& = \sum_{i=1}^m \left(\frac{f_i}{\sqrt{m}}\right)^2.
\end{align*}
The eigendecomposition takes the form
\[
(k_{ji}) = \sum_{l=1}^m \tilde{\lambda}^l \tilde{f}^l (\tilde{f}^l)'
\]
and for a single measured element we get
\[
k_{ij} = \sum_{l=1}^m \tilde{\lambda}^l \tilde{f}^l_i \tilde{f}^l_j.
\]

\noindent According to the spectral theorem \cite{werner}
the eigendecomposition of $k$ is
\[
k(x,y) = \sum_{l=1}^\infty \gamma^l \phi^l(x) \phi^l(y)
\]
where $\gamma^l$ and $\phi^l$ are the eigenvalues and eigenfunctions, respectively.
Since $K$ is a compact operator, $\gamma^l$ is a null sequence.
Thus, the sequence of operators $\tilde{K}_m$ with the kernel
$\tilde{k}_m(x,y) = \sum_{l=1}^m \gamma^l \phi^l(x) \phi^l(y)$
converges to $K$ in the operator norm for $m \to \infty$ \cite{werner}
and we can approximate
\begin{align*}
k(x,y) & \approx
 \sum_{l=1}^m \gamma^l \phi^l(x) \phi^l(y) \\
& = \sum_{l=1}^m 
 \int_{\Omega} k(x,z) \phi^l(z) d\mu(z)
 \frac{1}{\gamma^l}
 \int_{\Omega} k(y,z') \phi^l(z') d\mu(z'),
\end{align*}
where we assume that none of the $\gamma^l$ are zero.
Further, due to the "Perturbation of bounded operators" theorem,
the eigenvalues $\lambda^l$ converge to $\gamma^l$
and the corresponding eigenspaces converge in the operator norm
and we can approximate
\[
k(x,y) \approx \sum_{l=1}^m 
 \int_{\Omega} k(x,z) f^l(z) d\mu(z)
 \frac{1}{\lambda^l}
 \int_{\Omega} k(y,z') f^l(z') d\mu(z').
\]
Taking into account the Equation \eqref{eig_func} the above formula turns into
\begin{align*}
k(x,y) \approx & \sum_{l=1}^m 
 \int_{\Omega} k(x,z) \sum_{i=1}^m f^l_i 1_{\Omega_i}(z) d\mu(z)\\
\cdot \; &
 \frac{1}{\lambda^l}
 \int_{\Omega} k(y,z') \sum_{j=1}^m f^l_j 1_{\Omega_j}(z') d\mu(z') \\
 = & \sum_{l=1}^m 
 \sum_{i=1}^m f^l_i \int_{\Omega_i} k(x,z) d\mu(z)
 \frac{1}{\lambda^l}
 \sum_{j=1}^m f^l_j \int_{\Omega_j} k(y,z') d\mu(z') \\
 = & \sum_{i=1}^m \sum_{j=1}^m
 k_i(x)
\left( 
 \sum_{l=1}^m \frac{f^l_i}{\sqrt{m}}  \frac{1}{m \lambda^l} \frac{f^l_j}{\sqrt{m}}
\right)
 k_j(y)\\
 = & \sum_{i=1}^m \sum_{j=1}^m
 k_i(x)
 \left( k^{-1} \right)_{ij}
 k_j(y),
\end{align*}
where $k^{-1}$ is the pseudo-inverse of the matrix consisting of elements $k_{ij}$.
It is now clear, that after measuring $k_i(x)$ at $N$ places
and writing the above formula in matrix form,
we retain the original Nystr\"om approximation as in Equation \eqref{eq:Ny_equation}.

Note, that the approximation of $k(x,y)$ consists of two approximations.
The first one is the approximation of the rank of the matrix and the second one
is the approximation of the eigenfunctions and eigenvalues.
Although we don't know the exact eigenvalues and eigenfunctions
of kernel $k(x,y)$, the approximation is exact if the kernel has a rank $\le m$
\footnote{If the true rank is larger than $m$ the eigenvalues do not match the 
true once and errors occur - like with any other approach. However the presented
approach can also keep negative eigenvalues, given they are within the top $m$ eigenvalues.}.
This fact is known for the Nystr\"om approximation and can be validated
by simple matrix transformations. The reason is, that if the rank of a kernel is $m$
then it can be represented as an inner product in a pseudo-Euclidean space
and $m$ linearly independent landmarks build a basis which spans this space.
The position of any new point $x$ is then fully determined by $k(x,x_i)$,
with $x_i$ being the landmarks, so that all inner products between any points
are determined and the matrix $\mathbf{K}$ can be computed precisely.

The Nystr\"om approximation involves the computation of $\mathbf{K}_{N,m}$
and inversion of $\mathbf{K}_{m,m}$ with the corresponding complexities of
$\mathcal{O}(mN)$ and $\mathcal{O}(m^3)$, respectively.
The multiplication of both matrices as well as multiplication
of the approximated matrix with other matrices,
required for further processing and training,
has the complexity of $\mathcal{O}(m^2 N)$.
%The computation of the eigenvalue decomposition requires additional
%matrix multiplications and eigenvalue decompositions
%with the complexity $\mathcal{O}(m^2N+m^3)$.
Thus, the overall complexity of the Nystr\"om technique
is given by $\mathcal{O}(m^2N)$.

\section{Transformations of (dis-)similarities with linear costs}
The Nystr\"om approximation was proposed originally
to deal with large psd similarity matrices
with kernel approaches in mind by \cite{DBLP:conf/nips/WilliamsS00}.
To apply these techniques on indefinite similarity and dissimilarity matrices
additional transformations, as discussed in section \ref{sec:trafos}, are required.
Unfortunately, these transformations have quadratic or even cubic time complexity,
making the advantage gained by the Nystr\"om approximation pointless.
Since we can now apply the Nystr\"om technique on arbitrary symmetric matrices,
it is not only possible to approximate the dissimilarities directly,
but also to perform the transformations in linear time.
Thus, we can apply relational and kernel techniques
on similarities and dissimilarities including eigenvalue corrections if necessary.

\begin{figure*}
\centering
	\includegraphics[width=0.7\columnwidth]{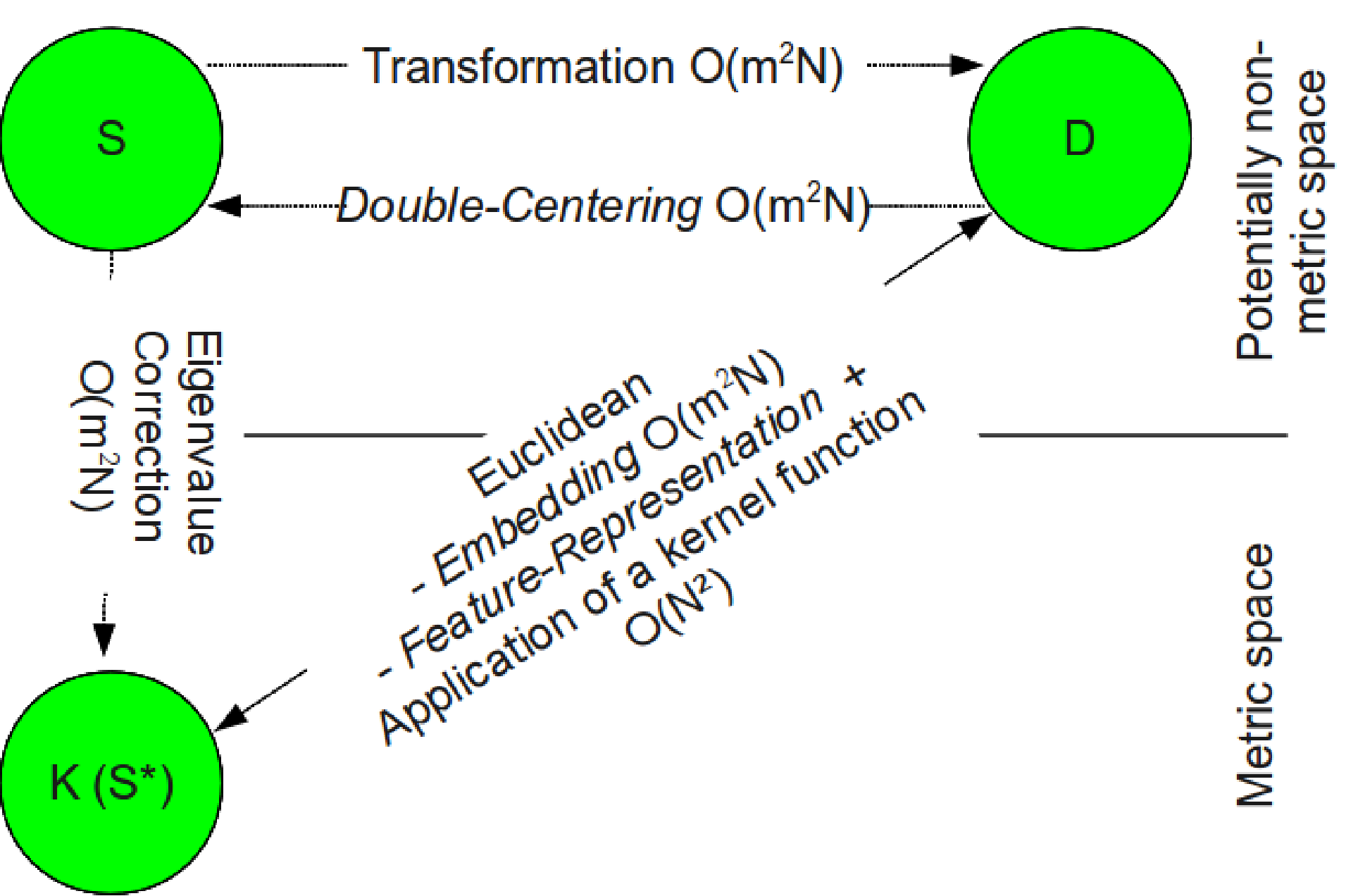}
	\caption{Updated schema from Figure \ref{fig:simdis_schema}
	using the discussed approximation.
	The costs are now substantially smaller, provided $m \ll N$.
	}
	 \label{fig:simdis_schema_ny}
\end{figure*}

In this section we will elaborate
how the transformations discussed in section \ref{sec:trafos}
can be done in linear time if applied for the Nystr\"om-approximated matrices.
The updated costs are shown on the Figure \ref{fig:simdis_schema_ny}.

\subsection{Transformation of dissimilarities and similarities into each other}
Given a dissimilarity matrix $\mathbf{D}$,
there are two ways to construct the approximated matrix $\mathbf{\hat{S}}$.
First, we can transform $\mathbf{D}$ to $\mathbf{S}$ using double centering
and then apply Nystr\"om approximation to $\mathbf{S}$.
Obviously, this approach has quadratic time complexity
due to the double centering step.
Second, we can approximate $\mathbf{D}$ to $\mathbf{\hat{D}}$ first
and then apply double centering.
As we will show in the following,
this transformation requires only linear computational time.

As mentioned before, from the dissimilarity matrix $\mathbf{D}$
we can compute the corresponding similiarity matrix using double centering.
This process is noted as  $\mathbf{S(D)}$ in the following:
\begin{equation*}\label{eq:double_centering}
	\mathbf{S(D)} = -\mathbf{J} \mathbf{D} \mathbf{J}/2 \\
\end{equation*}
where $\mathbf{J}=(\mathbf{I}-\mathbf{1}\mathbf{1}^\top/N)$
with identity matrix $\mathbf{I}$  and vector of ones $\mathbf{1}$.
%$\mathbf{S}$ is positive semi-definite if and only if $\mathbf{D}$ is Euclidean.
Expanding the right side of the equation we get
\begin{eqnarray*}
\mathbf{S(D)} &=& -\frac{1}{2} \mathbf{J} \mathbf{D} \mathbf{J} \nonumber \\
  &=& -\frac{1}{2} \left ( \left (\mathbf{I} - \frac{1}{N} \mathbf{1} \mathbf{1}^\top \right )
        \mathbf{D}
        \left (\mathbf{I} - \frac{1}{N} \mathbf{1} \mathbf{1}^\top \right ) \right )
        \nonumber \\
%  &=& -\frac{1}{2} \left (\mathbf{I} \mathbf{D} \mathbf{I}
%      -\frac{1}{N} ( \mathbf{1} \mathbf{1}^\top \mathbf{D} \mathbf{I}
%      +\mathbf{I} \mathbf{D}  \mathbf{1} \mathbf{1}^\top 
%      -\mathbf{1} \mathbf{1}^\top
%       \mathbf{D}
%       \frac{1}{N} \mathbf{1} \mathbf{1}^\top ) \right ) \nonumber \\
  &=& -\frac{1}{2} \left (\mathbf{D} 
      -\frac{1}{N} \mathbf{D} \mathbf{1} \mathbf{1}^\top
      -\frac{1}{N} \mathbf{1} \mathbf{1}^\top \mathbf{D}
      +\frac{1}{N^2} \mathbf{1} \mathbf{1}^\top
       \mathbf{D}
       \mathbf{1} \mathbf{1}^\top \right ). \nonumber \\
\end{eqnarray*}
Approximating $\mathbf{S(D)}$ requires computation of a linear part of each summand,
but still involves summation over the full matrix $\mathbf{D}$.

Alternatively, by approximating $\mathbf{D}$ first, we get
\begin{eqnarray}
&\mathbf{S} \overset{Ny}{\approx}& \mathbf{S(\hat{D})}
  = -\frac{1}{2} \left[ \mathbf{D}_{N,m} \cdot \mathbf{D}_{m,m}^{-1} \cdot \mathbf{D}_{m,N} -\frac{1}{N} \mathbf{D}_{N,m} \right. \label{eq:dis_to_sim} \\
  & & \left.\cdot (\mathbf{D}_{m,m}^{-1} \cdot (\mathbf{D}_{m,N} \mathbf{1} )) \mathbf{1}^\top -\frac{1}{N} \mathbf{1} ((\mathbf{1}^\top \mathbf{D}_{N,m})  \cdot \mathbf{D}_{m,m}^{-1}) \right.
\nonumber\\
  & & \left.  \cdot \mathbf{D}_{m,N}
	  +\frac{1}{N^2} \mathbf{1} (( \mathbf{1}^\top  \mathbf{D}_{N,m}) \cdot
	   \mathbf{D}_{m,m}^{-1} \cdot (\mathbf{D}_{m,N} \mathbf{1}))\mathbf{1}^\top
	   \right]. \nonumber
\end{eqnarray}
This equation can be rewritten for each entry of the matrix $\mathbf{S(\hat{D})}$
\begin{eqnarray*}
\hat{S}_{ij}(\mathbf{\hat{D}})
  &=& -\frac{1}{2} \Bigg[ \mathbf{D}_{i,m} \cdot \mathbf{D}_{m,m}^{-1} \cdot \mathbf{D}_{m,j} \\
  & &    -\frac{1}{N} \sum_k \mathbf{D}_{k,m} \cdot \mathbf{D}_{m,m}^{-1} \cdot
       \mathbf{D}_{m,j} \\
  & & \left.
      -\frac{1}{N} \sum_k \mathbf{D}_{i,m} \cdot \mathbf{D}_{m,m}^{-1} \cdot
       \mathbf{D}_{m,k}\right.\nonumber\\
  & &  \left.    +\frac{1}{N^2} \sum_{kl} \mathbf{D}_{k,m} \cdot
       \mathbf{D}_{m,m}^{-1} \cdot \mathbf{D}_{m,l}
      \right], \nonumber
\end{eqnarray*}
as well as for the sub-matrices $\mathbf{S}_{m,m}(\mathbf{\hat{D}})$ and $\mathbf{S}_{N,m}(\mathbf{\hat{D}})$,
in which we are interested for the Nystr\"om approximation 
\begin{eqnarray*}
\mathbf{S}_{m,m}(\mathbf{\hat{D}})
  &=& -\frac{1}{2} \left[ \mathbf{D}_{m,m}
      -\frac{1}{N} \mathbf{1} \cdot \sum_k \mathbf{D}_{k,m} \right. \\
  & & \left.
      -\frac{1}{N} \sum_k \mathbf{D}_{m,k} \cdot \mathbf{1}^\top \right.\nonumber\\
  &&  \left.     +\frac{1}{N^2} \mathbf{1} \cdot \sum_{kl} \mathbf{D}_{k,m} \cdot
       \mathbf{D}_{m,m}^{-1} \cdot \mathbf{D}_{m,l} \cdot \mathbf{1}^\top
      \right] \nonumber
\end{eqnarray*}
\begin{eqnarray*}
\mathbf{S}_{N,m}(\mathbf{\hat{D}})
  &=& -\frac{1}{2} \left[ \mathbf{D}_{N,m}
      -\frac{1}{N} \mathbf{1} \cdot \sum_k \mathbf{D}_{k,m} \right. \\
  & & \left.
      -\frac{1}{N} \sum_k \mathbf{D}_{N,m} \cdot \mathbf{D}_{m,m}^{-1} \cdot
       \mathbf{D}_{m,k} \cdot \mathbf{1}^\top\right. \nonumber\\
  & & \left.    +\frac{1}{N^2} \mathbf{1} \cdot \sum_{kl} \mathbf{D}_{k,m} \cdot
       \mathbf{D}_{m,m}^{-1} \cdot \mathbf{D}_{m,l} \cdot \mathbf{1}^\top
      \right]. \nonumber
\end{eqnarray*}
Now the matrix $\mathbf{S(\hat{D})}$ can be approximated via the matrix
$\mathbf{\hat{S}(\hat{D})}$ using the matrices
$\mathbf{S}_{m,m}(\mathbf{\hat{D}})$ and $\mathbf{S}_{N,m}(\mathbf{\hat{D}})$.
This requires only a linear part of $\mathbf{D}$ and involves linear computation time.

Comparing this approach to the quadratic computation of $\mathbf{S}_{N,m}$,
we see, that the first three summands are identical
and only the forth summand is different.
This term involves summation over the full dissimilarity matrix
and, depending on the approximation quality of $\mathbf{\hat{D}}$, might vary.
The deviation is added to each pairwise similarity
resulting in a non-linear transformation of the data.
If $m$ corresponds to the rank of $\mathbf{D}$
then double centering is exact
and no information loss occurs during the approximation.
Otherwise, the information loss increases with smaller $m$
for both approaches
and the error is made by approximating $\mathbf{S}$ in the first case
and by approximating $\mathbf{D}$ in the second case.
If the Nystr\"om approximation is feasible for a given data set,
then the second approach allows
to perform the transformation in linear instead of quadratic time.

It should be mentioned that a similar transformation is possible with
the landmark multidimensional scaling (L-MDS) \cite{DBLP:conf/nips/SilvaT02}
which is widely known in the visualization community and typically used to 
embed data into a low $2-3$ dimensional space. Embeddings to higher dimensions
are possible but not considered, in general.
The idea of L-MDS is to sample a small amount $m$ of points, the so called landmarks,
compute the corresponding dissimilarity matrix
followed by a double centering on this matrix.
Finally the data are projected to a low dimensional space
using an eigenvalue decomposition.
The remaining points can then be projected into the same space,
taking into account the distances to the landmarks, and applying a triangulation.
From this vectorial representation of the data
one can easily retrieve the similarity
matrix as a scalar product between the points.

It was shown, that L-MDS is also a Nystr\"om technique by \cite{Platt:2005},
but compared to our proposed approach in Equation \eqref{eq:dis_to_sim}
L-MDS makes not only an error in the forth summand, but also in the second and the third.
Additionally, and more importantly, by projecting into \emph{Euclidean space}
it makes an implicit clipping of the eigenvalues. 
As discussed above and will be shown later,
this might disturb data significantly, leading to qualitatively worse results.
Thus, our proposed method can be seen as a generalization of L-MDS
and should be used instead.

Similarly to the transformation from $\mathbf{D}$ to $\mathbf{\hat{S}}$,
there are two ways to transform $\mathbf{S}$ to $\mathbf{\hat{D}}$.
First, transform the full matrix $\mathbf{S}$ to $\mathbf{D}$
using $D_{ij} = S_{ii} + S_{jj} - 2 S_{ij}$
and then apply the Nystr\"om approximation
\begin{equation}
\mathbf{\hat{D}} = \mathbf{D}_{N,m}
\mathbf{D}_{m,m}^{-1}
\mathbf{D}_{N,m}^\top.
\label{eq:dis_corr_app}
\end{equation}
Second, approximate $\mathbf{S}$ with $\mathbf{\hat{S}}$
and then transform it to $\mathbf{\hat{D}}$.
The first approach requires quadratic time,
since it transforms the full matrix.
In the second approach only $\mathbf{D}_{N,m}$ is computed,
thus making it linear in time and memory.
Obviously, both approaches produce the same results,
but the second one is significantly faster.
The reason is, that for the computation of $\mathbf{\hat{D}}$
only the matrix $\mathbf{D}_{N,m}$ is required
and it is not necessary to compute the rest of $\mathbf{D}$.

\subsection{Eigenvalue correction}
For non-Euclidean data,
the corresponding similarity matrix is indefinite.
We would like to make the data Euclidean
in order to avoid convergence issues,
or to be able to use kernel methods.
A strategy to obtain a valid kernel matrix from similarities
is to apply an eigenvalue correction as discussed in section \ref{sec:trafos_corr}. 
This however can be prohibitive for large matrices, since to correct the whole
eigenvalue spectrum, the whole eigenvalue decomposition is needed,
which has $\mathcal{O}(N^3)$ complexity.
The Nystr\"om approximation can again decrease computational costs dramatically.
Since we can now apply the approximation on an arbitrary symmetric matrix,
we can make the correction afterwards,
reducing the complexity to a linear one, as we will show now.

Given non-metric dissimilarities $\mathbf{D}$,
we can first approximate them and then convert to approximated similarities $\mathbf{\hat{S}}(\mathbf{\hat{D}})$
using the Equation \eqref{eq:dis_to_sim}.
For similarities $\mathbf{\hat{S}}$
given directly or obtained from $\mathbf{\hat{S}}(\mathbf{\hat{D}})$,
we need to compute the eigenvalue decomposition in linear time.
As we have shown in the section \ref{sec:eval_decomp},
it is possible to compute the exact eigenvalue decomposition
of a Nystr\"om-approximated psd matrix in linear time, given
the corresponding similarity matrix has indeed rank $m$.
Since $\mathbf{\hat{S}}$ is indefinite,
we can not apply the above technique directly.
Instead, since in a squared matrix the eigenvectors stay the same,
we first compute %$\mathbf{\hat{S}}^2$ and determine its eigenvectors
\begin{align*}
\mathbf{\hat{S}}^2 & = \mathbf{S}_{N,m} \mathbf{S}^{-1}_{m,m} \left( \mathbf{S}_{m,N}
	\cdot \mathbf{S}_{N,m} \right) \mathbf{S}^{-1}_{m,m} \mathbf{S}_{m,N}\\
& = \mathbf{S}_{N,m} \mathbf{\tilde{S}}_{m,m} \mathbf{S}_{N,m}^\top\\
& = \mathbf{C} \mathbf{\tilde{A}} \mathbf{C}^\top.
\end{align*}
The resulting matrix can be computed in linear time and is psd.
This means, we can determine its eigenvalue decomposition
as described in section \ref{sec:eval_decomp}:
\[
\mathbf{\hat{S}}^2 = \mathbf{C} \mathbf{\tilde{A}} \mathbf{C}^\top,
\]
where $\mathbf{\tilde{A}}$ are the eigenvalues of $\mathbf{\hat{S}}^2$
and $\mathbf{C}$ are the eigenvectors
of both $\mathbf{\hat{S}}^2$ and $\mathbf{\hat{S}}$.

Using the eigenvectors $\mathbf{C}$, the eigenvalues $\mathbf{A}$
of $\mathbf{\hat{S}}=\mathbf{C}\mathbf{A}\mathbf{C}^\top$
can be retrieved via

$\mathbf{A}=\mathbf{C}^\top \mathbf{\hat{S}} \mathbf{C}$.
Then we can correct the eigenvalues $\mathbf{A}$
by some technique as discussed in section \ref{sec:trafos_corr} to $\mathbf{A}^*$.
The corrected approximated matrix $\mathbf{\hat{S}}^*$ is then simply
\begin{equation}
\label{eq:sim_corr_app}
\mathbf{\hat{S}}^* = \mathbf{C} \mathbf{A}^* \mathbf{C}^\top.
\end{equation}
%\begin{equation}
%\label{eq:sim_corr_app}
%\mathbf{\hat{S}}^* = 
%	\mathbf{S}_{N,m} \mathbf{U}
%	\left( \mathbf{\Lambda}^* \right)^{-1} 
%	\mathbf{U}^\top \mathbf{S}_{N,m}^\top.
%\end{equation}
Thus, using a low rank representation of a similarity matrix
we can compute its eigenvalue decomposition
and perform eigenvalue correction in linear time.
If it is desirable to work with the corrected dissimilarities,
then using the Equation \eqref{eq:dis_corr_app}, it is possible to transform
the corrected similarity matrix $\mathbf{\hat{S}}^*$ back to dissimilarities
resulting in the corrected and approximated matrix $\mathbf{\hat{D}}^*$.

\subsection{Out-of-sample extension}
Usually models are learned by a training set
and we expect them to generalize well on the new unseen data, or the test set.
In such cases we need to provide an out-of-sample extension,
i.e.\ a way to apply the model on the new data.
This might be a problem for the techniques dealing with (dis-)similarities.
For example, in proxy approaches the out of sample extension is in general
handled by solving another costly optimization problem \cite{DBLP:conf/icml/ChenGR09,Lu30082005}.
If the matrices are corrected, we need to correct the new (dis-)similarities
as well to get consistent results. Fortunately this can be easily done in the
Nystr\"om framework. 

If we compare the Equations \eqref{eq:Ny_equation} and \eqref{eq:sim_corr_app}
we see that the correction is performed
on a different decomposition of $\mathbf{\hat{S}}$, i.e.:
\begin{equation}
\label{eq:appr_ooe}
\mathbf{S}_{N,m} \mathbf{S}_{m,m} \mathbf{S}_{N,m}^\top =
\mathbf{\hat{S}} = \mathbf{C} \mathbf{A} \mathbf{C}^\top.
\end{equation}
If we correct $\mathbf{A}$ it is not clear what happens on the left side
of the above equation.
Therefore, to compute the out-of-sample extension
we need to find a simple transformation
from one decomposition to the other.
Taking a linear part $\mathbf{\hat{S}}_{N,m}$ from the equation \ref{eq:appr_ooe}
we get
\[
\mathbf{S}_{N,m} =
\mathbf{C}_{N,m} \mathbf{A} \mathbf{C}_{m,m}^\top,
\]
which leads after simple transformation to
\[
\mathbf{C}_{N,m} = \mathbf{S}_{N,m}
\left( \mathbf{A} \mathbf{C}_{m,m}^\top \right)^{-1}.
\]
Plugging the above formula into Equation \eqref{eq:sim_corr_app} we get
\begin{align*}
\mathbf{\hat{S}}^* & =
\mathbf{S}_{N,m} \left( \mathbf{A} \mathbf{C}_{m,m}^\top \right)^{-1}
\mathbf{A}^*
\left(\left( \mathbf{A} \mathbf{C}_{m,m}^\top \right)^{-1}\right)^\top \mathbf{S}_{N,m}^\top\\
& = \mathbf{S}_{N,m} (\mathbf{C}_{m,m}^\top)^{-1}\mathbf{A}^{-1}
\mathbf{A}^*
\mathbf{A}^{-1}\mathbf{C}_{m,m}^{-1}\mathbf{S}_{N,m}^\top\\
& = \mathbf{S}_{N,m} (\mathbf{C}_{m,m}^\top)^{-1}
(\mathbf{A}^*)^{-1}
\mathbf{C}_{m,m}^{-1}\mathbf{S}_{N,m}^\top\\
& = \mathbf{S}_{N,m}
\left(\mathbf{C}_{m,m}\mathbf{A}^*\mathbf{C}_{m,m}^\top\right)^{-1}
\mathbf{S}_{N,m}^\top
\end{align*}
and we see that we simply need to extend the matrix $\mathbf{S}_{N,m}$
by uncorrected similarities between the new points and the landmarks to obtain
the full approximated and \emph{corrected} similarity matrix,
which then can be used by the algorithms to compute the out-of-sample extension.
The same approach can be applied to the dissimilarity matrices.
Here we first need to transform the new dissimilarities to similarities
using Equation \eqref{eq:dis_to_sim}, correct them and then
transform back to dissimilarities.
%By examining the equations \ref{eq:sim_corr_app} and
%\ref{eq:dis_corr_app} we see, that we simply need to extend the matrices
%$\mathbf{D}_{N,m}$ or $\mathbf{S}_{N,m}$, respectively, by uncorrected
%(dis)similarities between the new points and the landmarks to obtain
%the full approximated and \emph{corrected} (dis)similarity matrices,
%which then can be used by the algorithms to compute the out of sample extension.

In \cite{DBLP:journals/jmlr/ChenGGRC09} a similar approach is taken.
First, the whole similarity matrix
is corrected by means of a projection matrix. Then this projection matrix is
applied to the new data, so that the corrected similarity between old and new
data can be computed. This technique is in fact the Nystr\"om approximation,
where the whole similarity matrix $\mathbf{S}$ is treated as the approximation
matrix $\mathbf{S}_{m,m}$ and the old data, together with the new data build the
matrix $\mathbf{S}_{N,m}$. Rewriting this in the Nystr\"om framework
makes it clear and more obvious, without the need to compute the projection matrix and
with an additional possibility to compute the similarities between the new points.
%In Figure \ref{fig:simdis_schema_ny} we depict schematically the new situation 
%for similarity and dissimilarity data incorporating the proposed approach. 

\subsection{Proof of concept}
We close this section by a small experiment on the ball dataset as proposed in \cite{DBLP:conf/sspr/DuinP10}. 
It is an artificial dataset based on the surface distances of randomly positioned balls of two classes having a slightly different radius.
The dataset is non-Euclidean with substantial information encoded in the negative part of the eigenspectrum. 
We generated the data with $300$ samples per class leading to an $N \times N$
dissimilarity matrix $\mathbf{D}$, with $N=600$.
Now the data have been processed in four different ways
to obtain a valid kernel matrix $\mathbf{S}$.
First encoding, denoted as $SIM1$, was constructed
by converting $\mathbf{D}$ to $\mathbf{S}$
with double centering and computing the full eigenvalue decomposition.
The negative eigenvalues were then corrected by flipping.
This approach, which we will refer to as the {\bf standard approach} in the following,
has a complexity of $\mathcal{O}(N^3)$. 

Further, we generated an approximated similarity matrix $\mathbf{\hat{S}}^*$ by using the proposed approach, flipping in the eigenvalue correction
and $10$ landmarks for the Nystr\"om approximation. This dataset is denoted as $SIM2$ and was obtained with a complexity of $\mathcal{O}(m^2N)$.
The third dataset $SIM3$ was obtained in the same way but the eigenvalues were clipped. The dataset $SIM4$ was obtained using
landmark MDS with the same landmarks as for $SIM2$ and $SIM3$.
The data are processed by a Support Vector Machine in a $10$-fold
crossvalidation. The results on the test sets are shown in the Table \ref{tab:sim}. 

\begin{table*}
\centering
\caption{\label{tab:sim} Test set results of a 10-fold SVM run on the ball dataset using the different encodings.}

\begin{tabular*}{\textwidth}{@{\extracolsep{\fill}}l|c|c|c|c|c}\hline
	 		& $SIM1$  		& $SIM2$ 			& $SIM3$			& $SIM4$ \\\hline\hline
Test-Accuracy	& $100\pm0$	& $88.83\pm3.15 $	& $51.50\pm6.64 $ 	& $50.67\pm3.94$
\end{tabular*}
\end{table*}				

As mentioned, the data contain substantial information in the negative fraction of the eigenspectrum, accordingly one may
expect that these eigenvalues should not be removed.
This is also reflected in the results. L-MDS removed the negative eigenvalues
and the classification model based on these data shows random prediction accuracy.
The SIM3 encoding is a bit better. 
Also in this case the negative eigenvalues are removed but the limited amount of class separation information, encoded 
in the positive fraction was better preserved, probably due to the different calculation of the matrix $\mathbf{\hat{S}}_{mm}$.
The SIM2 data used the flipping strategy and shows already quite good prediction accuracy, taking into account that the
kernel matrix is only approximated by $10$ landmarks and the relevant (original negative) eigenvalues are of small magnitude.

As a last point it should be mentioned that corrections like clipping, flipping and their effect
on the data representation are still under discussion and considered to be not always optimal \cite{Pekalska2005a}. 
%Accordingly, it is still best to avoid such transformations and focus
%either on the appropriate methods, less sensitive to such effects, e.g. dissimilarity learners for dissimilarity data or to use (dis-)similarity
%measures which imply a metric space. 
Additionally the selection of landmark points is discussed in \cite{DBLP:journals/tnn/ZhangK10a,DBLP:journals/jmlr/KumarMT12}
%for more complex data sets were an only random selection of landmarks maybe not sufficient to cover the whole data properties 
Further, for very large data sets (e.g. some 100 million points) the Nystr\"om approximation may still be 
too costly and some other strategies have to be found as suggested in \cite{DBLP:conf/icml/LiKL10}.

\section{Experiments}
We now apply the priorly derived approach to six non-metric dissimilarity and similarity data and show the effectiveness for a classification task.
The considered data are  (1)  the imbalanced SwissProt similarity data as described in \cite{mediansom} consisting of protein sequence alignment
scores, (2) the balanced chromosome dissimilarity data taken from \cite{neuhaus} with scores of aligned gray value images, (3) the imbalanced proteom dissimilarity 
data set  from \cite{PrTools:2012:Online}, (4) the balanced Zongker digit dissimilarity data from \cite{PrTools:2012:Online,Jain19971386} which
is based on deformable template matchings of 2000 handwritten NIST digits. Further the balanced Delft gestures data base (DS5) 
taken from \cite{PrTools:2012:Online} and the WoodyPlants50 (Woody) (DS6) from the same source is used. 
DS5 represents a sign-language interpretation problem with dissimilarities computed 
using a dynamic time warping procedure  on the sequence of positions \cite{Lichtenauer20082040}. 
The DS6 dataset contains of shape dissimilarities between leaves collected in a study on woody plants \cite{DBLP:journals/pami/LingJ07}.
Further details about the data can be found in Table \ref{tab:datasets}.

\begin{table*}[ht]
\begin{center}
\caption{\label{tab:datasets} Overview of the considered datasets and their properties.}
%\footnotesize
\begin{tabular*}{\textwidth}{@{\extracolsep{\fill}}l|c|c|c|c}\hline
Data set 	&	Name	 	&	\# samples & \# classes & Signature  \\\hline\hline
	DS1	&	SwissProt 		&	10988	 & 30		    & [8488,2500,0]\\
	DS2	&	Chromosom	&	4200		 & 21	             & [2258,1899,43]\\
	DS3	&	Proteom		& 	2604		 & 53            	   & [1502,682,420]\\
	DS4	&	Zongker		& 	2000		 & 10          	   & [961,1038,1]\\
	DS5	&	Delft		& 	1500		 & 20          	   & [963,536,1]\\
	DS6	&	Woody		& 	791		 	& 14          	 	   & [602,188,1]\\
\end{tabular*}
\end{center}
\end{table*}

All datasets are non-metric, multiclass and contain a large number of objects, such that a regular eigenvalue correction
with a prior double centering for dissimilarity data, as discussed before, is already very costly but can still be calculated to get comparative results. 

\subsection{Classification performance}
\begin{figure*}
\caption{Runtime analysis of the proposed vs the standard approach for the larger considered dissimilarity data
sets. All eigenvalues of the data sets have been processed by flipping. }\label{fig:diss_rt}
\begin{center}
	\subfigure[Chromosom]{\includegraphics[width=0.49\textwidth]{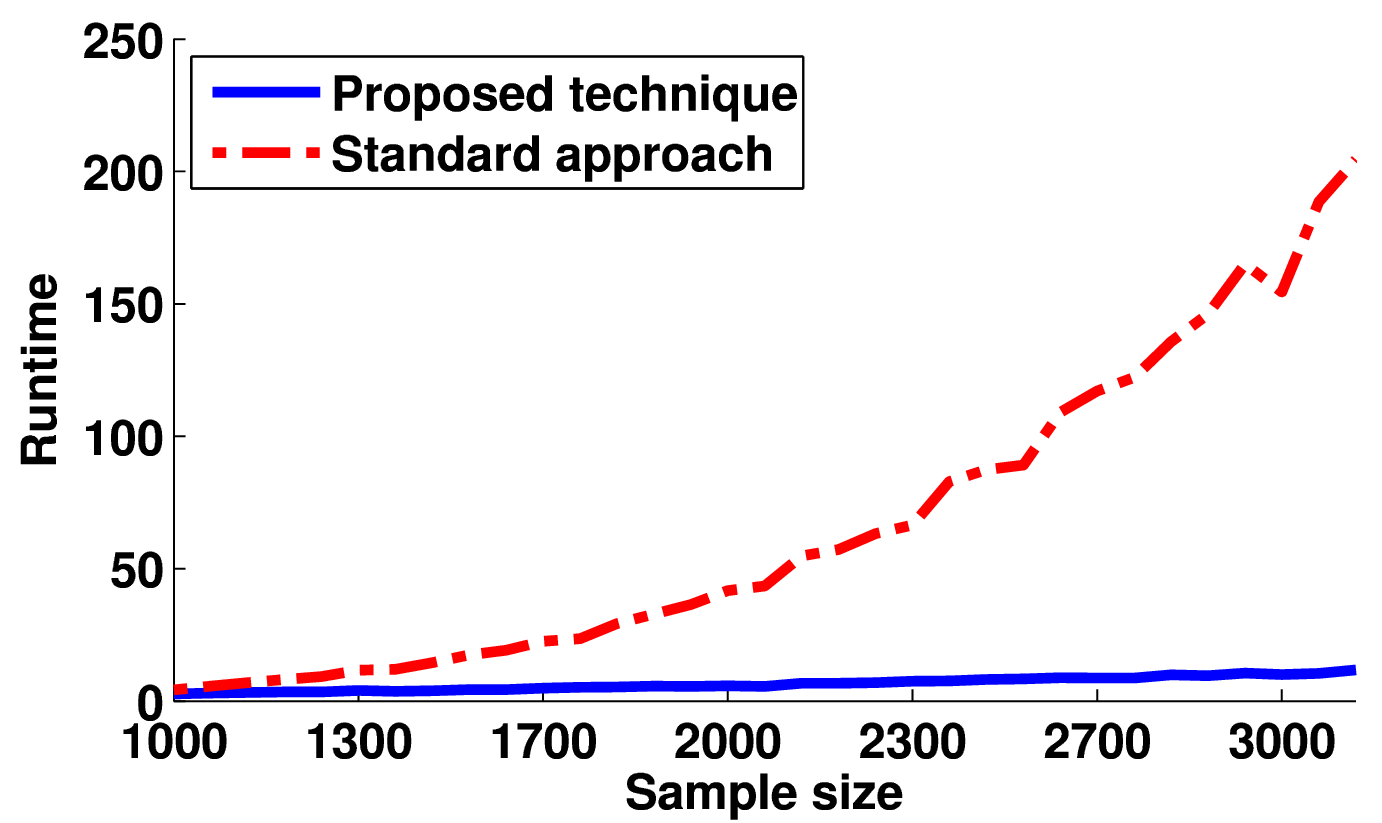}}
	\subfigure[Delft gestures]{\includegraphics[width=0.49\textwidth]{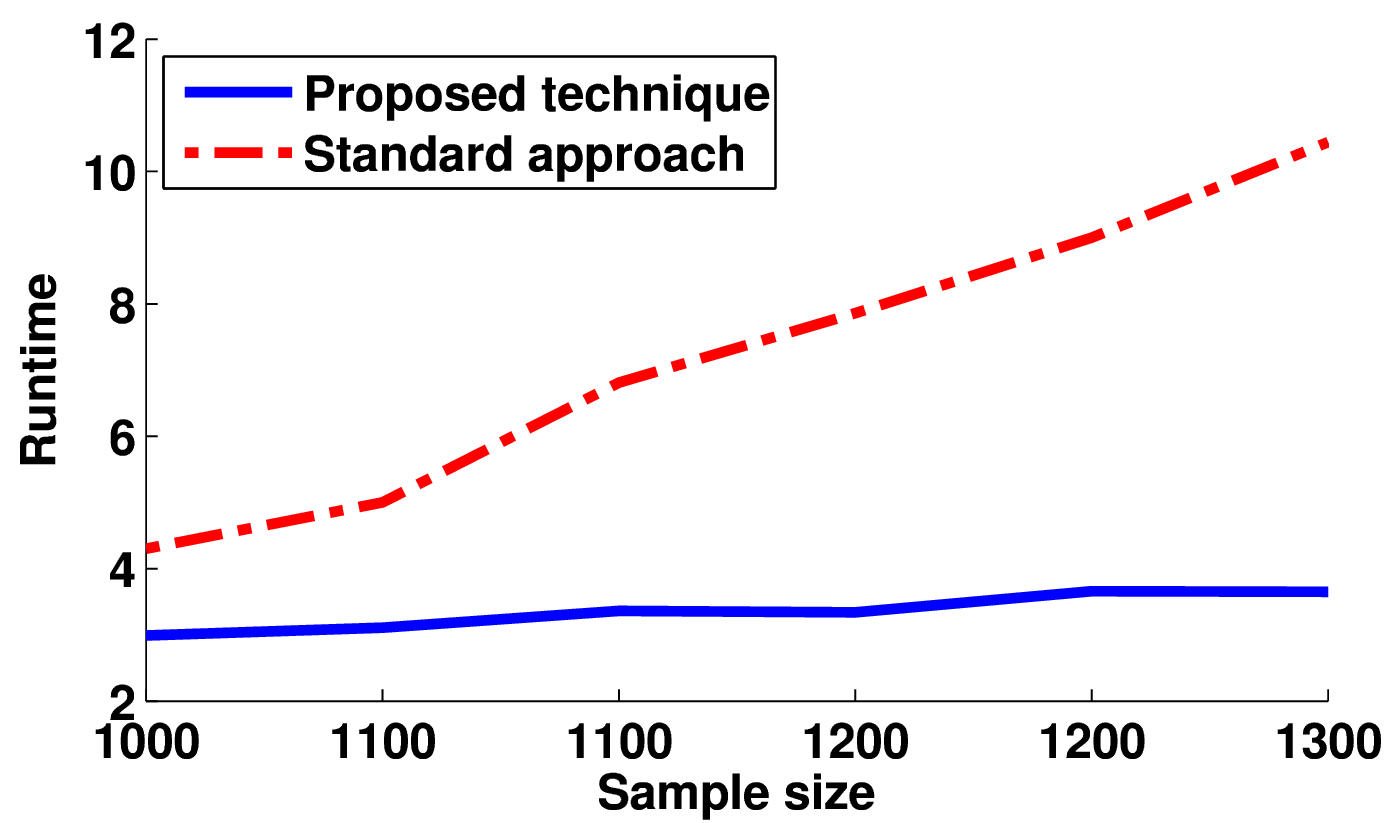}}\\
	\subfigure[Proteom]{\includegraphics[width=0.49\textwidth]{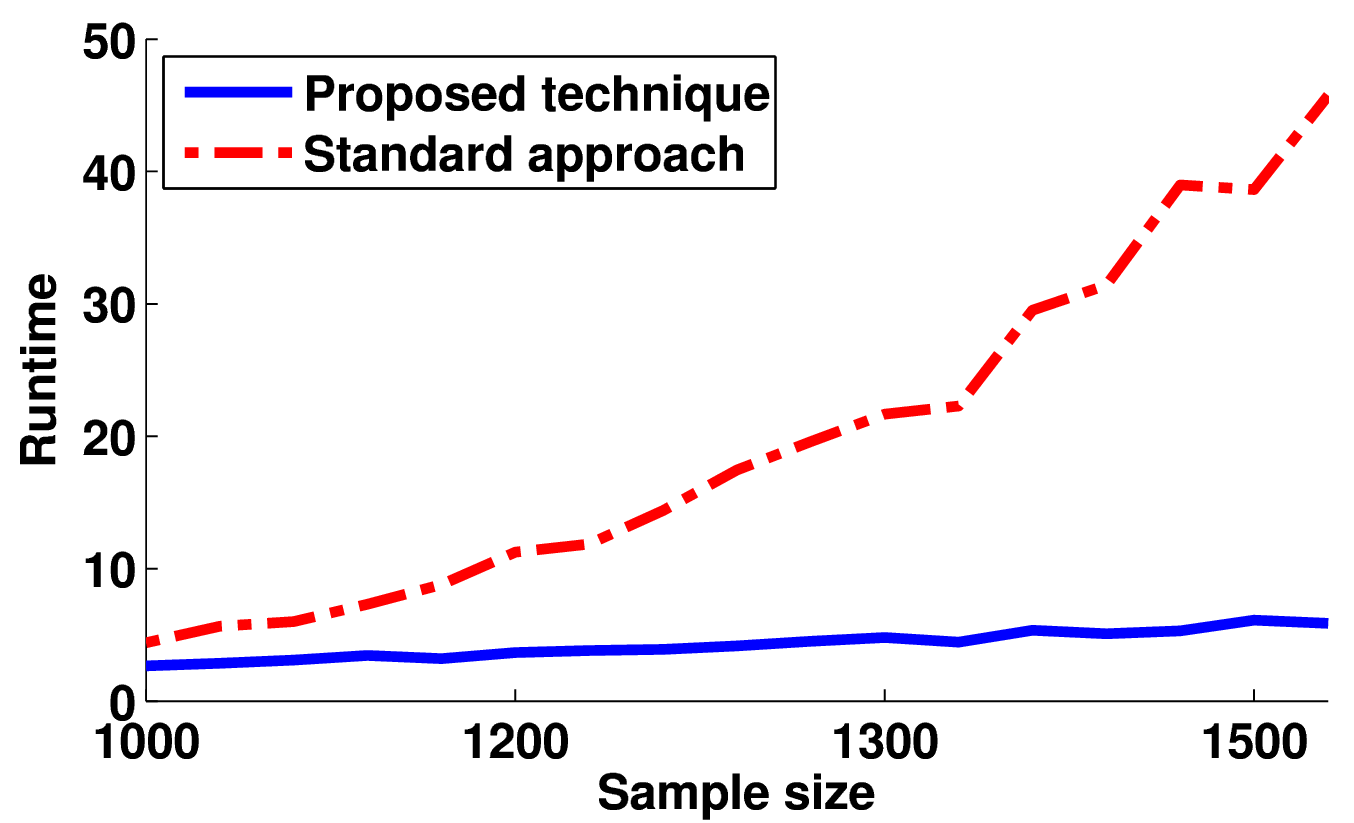}}
	\subfigure[Zongker]{\includegraphics[width=0.49\textwidth]{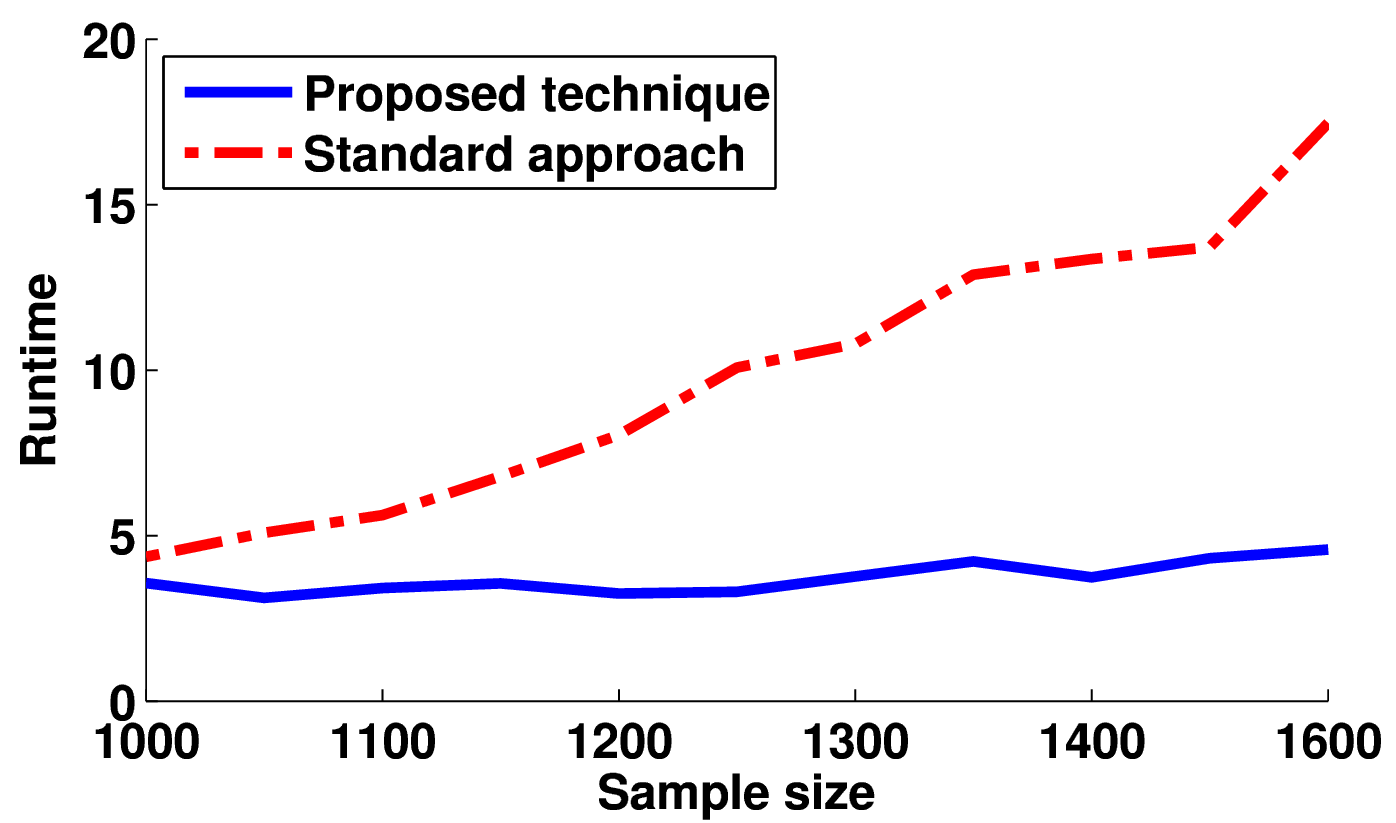}}
	\subfigure[SwissProt]{\includegraphics[width=0.49\textwidth]{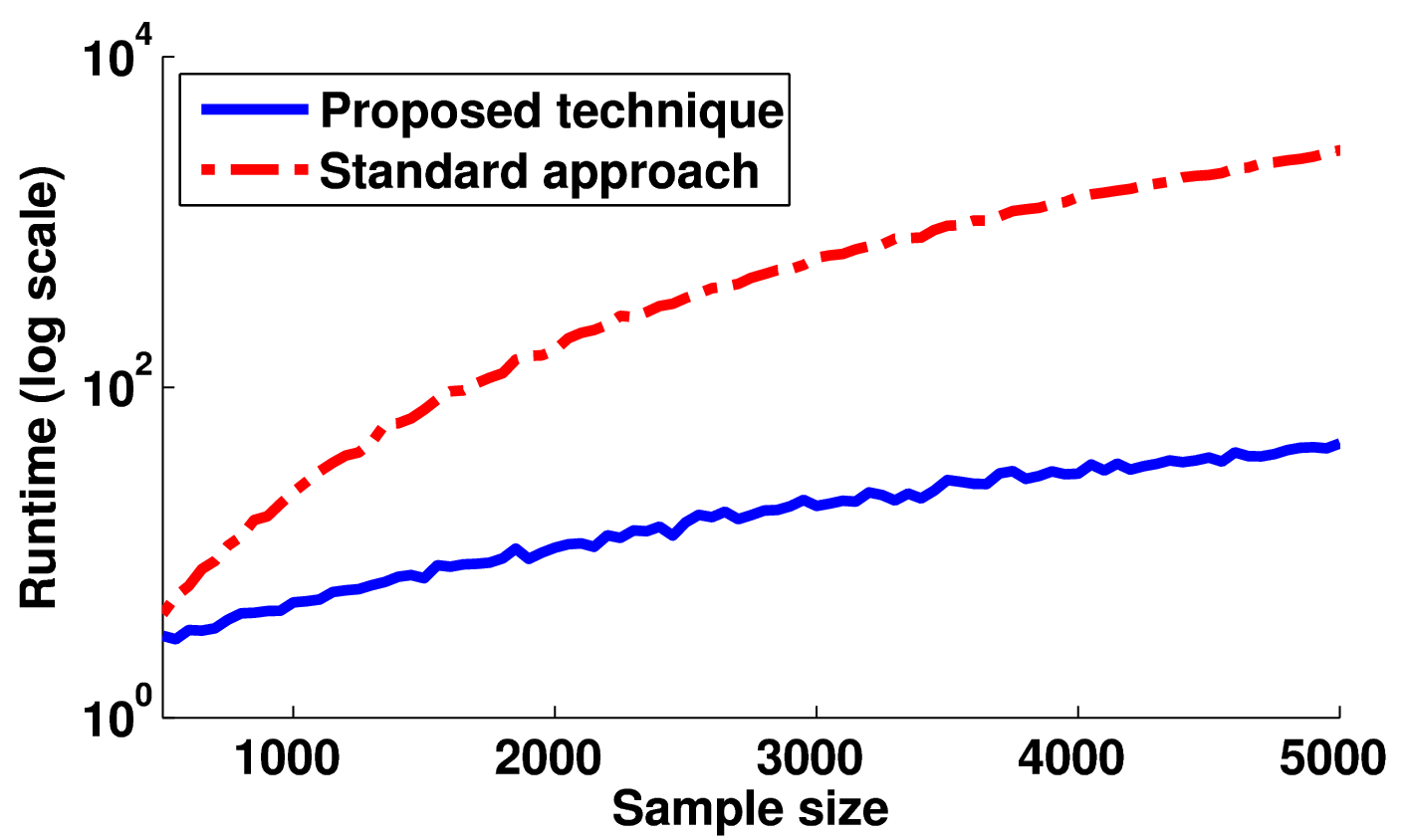}}
\end{center}
\end{figure*}

The data are analyzed in various ways, employing either the clipping eigenvalue correction, the flipping eigenvalue correction, or by not-correcting the eigenvalues
\footnote{
%Clipping and flipping were found similar effective, with a little advantage for flipping. With flipping the information of the negative-eigenvalues is at least somewhat kept  in the data representation so we focus on this representation. 
Shift correction was found to have a negative impact on the model as already discussed in \cite{DBLP:journals/jmlr/ChenGGRC09}.}. 
To be effective for the large number of objects we also apply the Nystr\"om approximation as discussed
before using $10, 50, 100$ and all points as landmarks. If the data have high rank $>100$, they are potentially not well suited for approximations
and approximation errors are unavoidable.
Landmarks have been selected randomly from the data. Other sampling strategies have been discussed in \cite{DBLP:journals/jmlr/FarahatGK11,DBLP:journals/tnn/ZhangK10a,DBLP:conf/icml/SiHD14}, 
however with additional meta parameters, which we would like to avoid for clarity of the proposed approach. Also the impact of the Nystr\"om approximation with respect
to kernel methods has been discussed recently in \cite{DBLP:journals/jmlr/CortesMT10}, 
but this is out of the focus of the presented approach.
For comparison we also show the results as obtained by using Landmark-MDS, which naturally applies a clipping and, as mentioned before,
makes various simplifications in the conversion step, which can lead to inaccuracies in the data representation.
\begin{figure}[p]
	\centering
	\includegraphics[width=0.99\textwidth]{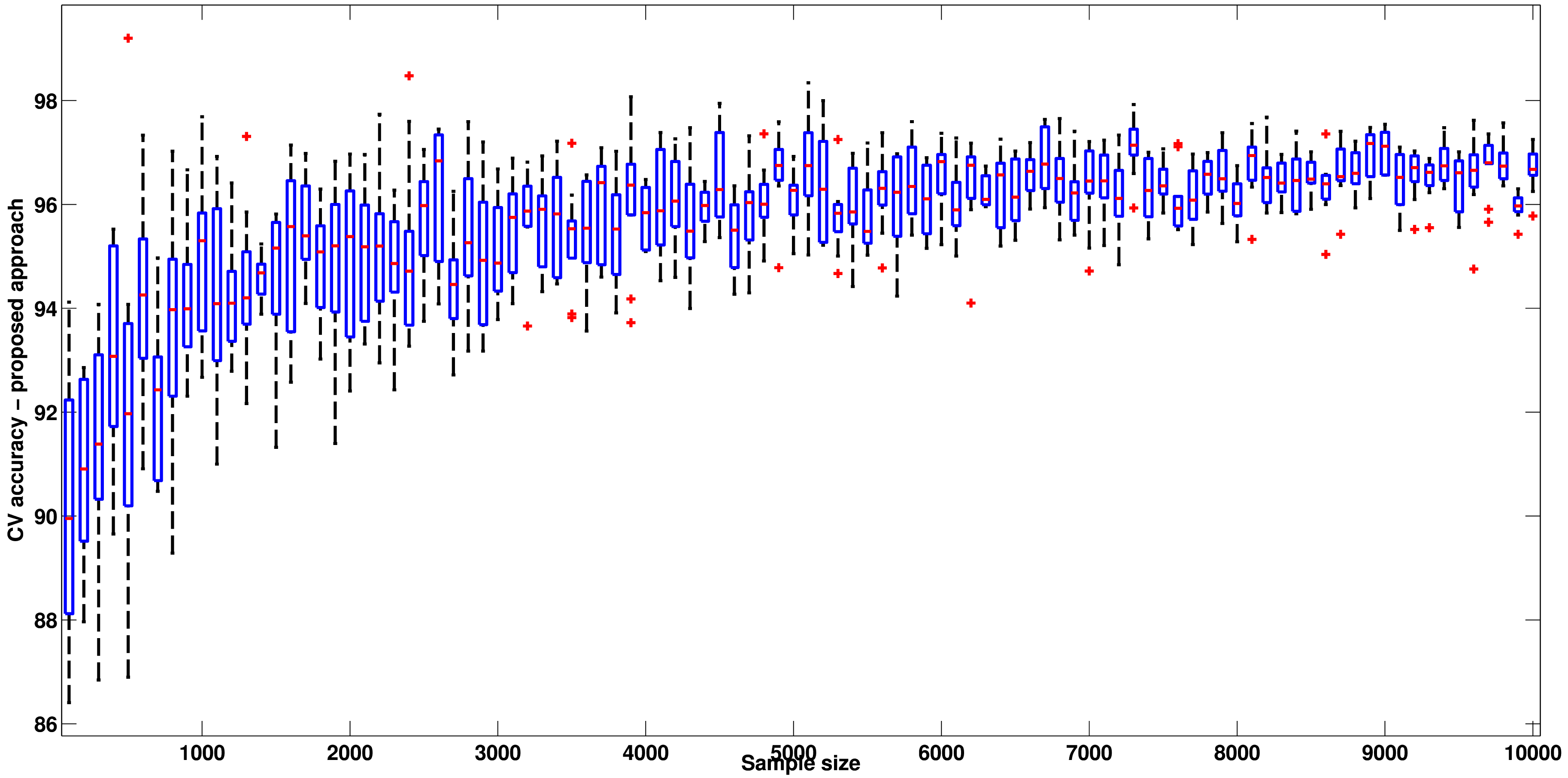}\\%\quad%\\
	\includegraphics[width=0.99\textwidth]{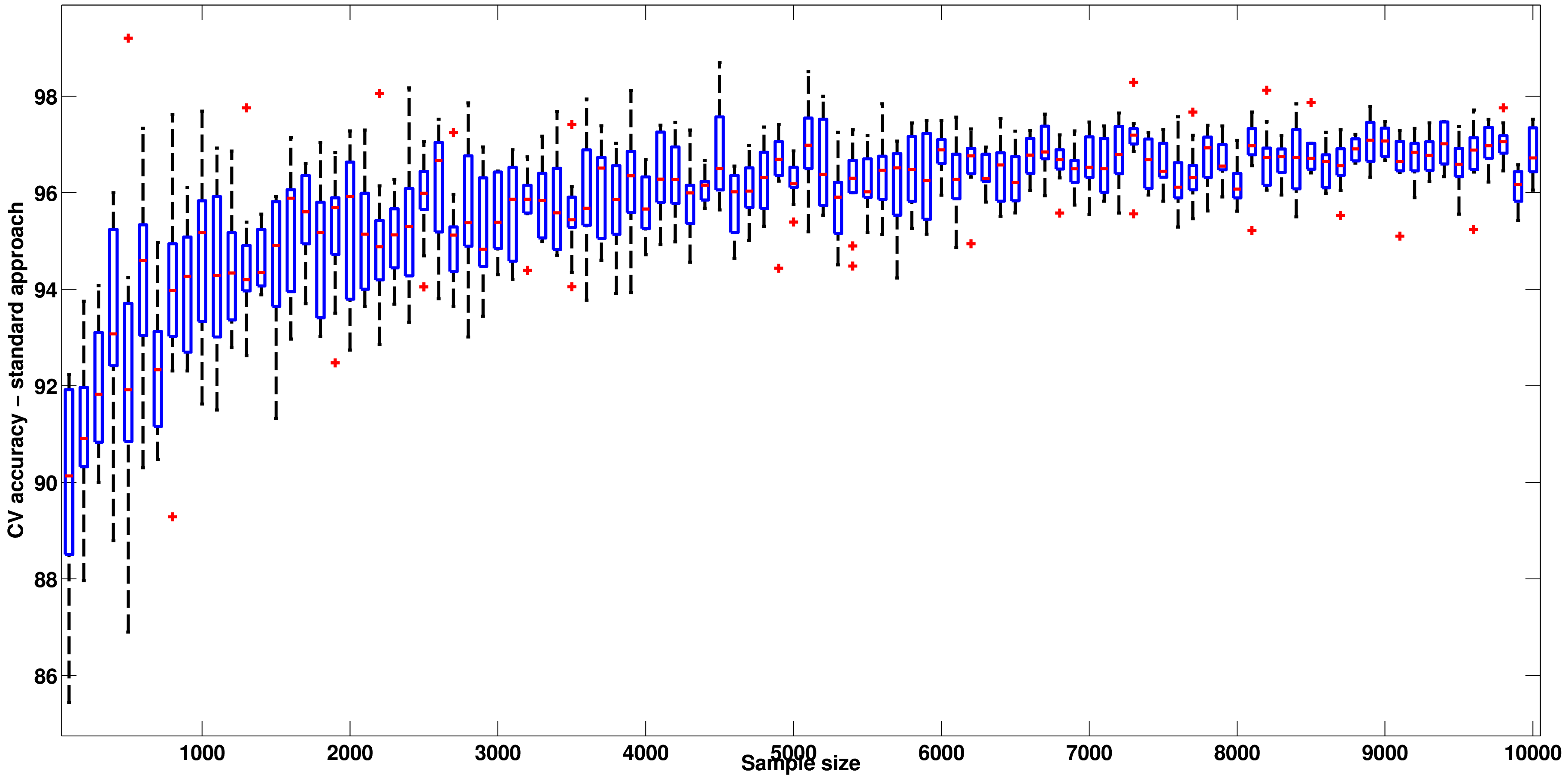}%\\
	\caption{Top: box-plots of the classification performance for different sample sizes of DS1 using the proposed approach with $500$ landmarks.
         Bottom: The same experiment but with the standard approach. Obviously our approach does not sacrifice performance for computational speed.}
         \label{fig:diss_rt_swiss}
\end{figure}
The prediction accuracies of a $10$-fold crossvalidation for $m=\{10,50,100\}$ are shown in Table \ref{tab:comparison_10}-\ref{tab:comparison_100}.
The influence of $N$ with respect to a fixed number of landmarks is studied in the experiment shown in Figure \ref{fig:diss_rt_swiss}.
A runtime analysis, comparing to the standard approach, is shown in Figure \ref{fig:diss_rt}. The results of the standard approach where no approximations are used but only eigenvalue corrections on the full matrix 
are provided in Table \ref{tab:comparison_full}. We also provide results for the dissimilarity-space representation using a linear and an elm kernel \cite{DBLP:journals/ijon/FrenayV11} in Table \ref{tab:comparison_diss_space}.
As mentioned before this representation does not need any approximations or eigenvalue corrections but the out of sample extension is
costly if many landmarks are chosen, or the selection of the landmarks has to be optimized using e.g. a wrapper approach \cite{DBLP:journals/pr/PekalskaDP06}.
Here we use all points as landmarks to simplify the evaluation.

\begin{table}
\begin{center}
\caption{\label{tab:comparison_10} Signature and average test set accuracy for SwissProt (DS1), Chromosome (DS2), Proteom (DS3), Zongker (DS4), Delft gestures (DS5), Woody (DS6)
	using a Nystr\"om approximation with $10, 50, 100, full$ landmarks and no, clip or flip eigenvalue correction.}
\footnotesize
\begin{tabular*}{1\textwidth}{@{\extracolsep{\fill}}|c|c|c|c|c}\hline
		& 		$10$ / Clip				& 	$10$ / Flip       			& 	$10$ / No			& 	$10$ L-MDS   								\\\hline\hline
 \tiny{DS1}& 	 \footnotesize{[9, 0, 10979]}		& \footnotesize{[10, 0, 10978]} 	&\footnotesize{[9, 1, 10978]}	&\\
		&  	 $30.67\pm5.07$*				& $\bf 31.65\pm 5.41	$*		&	$5.93 \pm 5.23$				&	$26.47 \pm 6.27$\\
 \tiny{DS2}&	\footnotesize{[9, 0 ,4191]}		& \footnotesize{[10, 0, 4190]} 	&\footnotesize{[9,1, 4190]}			&\\
		&   	 $67.61\pm6.49$				& $\bf 74.83\pm 3.23$*			&$ 18.79 \pm 14.08$			&$ 67.086\pm 6.09$	\\
\tiny{DS3}& 	\footnotesize{[9, 0 ,2595]}		& \footnotesize{[10, 0, 2594]} 	&\footnotesize{[9, 1, 2594]}		&\\
		& 	 $59.33\pm 6.87$*				& $\bf 62.43\pm 7.30	$*		&	$ 2.52\pm2.33$				&$ 56.74\pm6.26$	\\
\tiny{DS4}&	\footnotesize{[8, 0 ,1992]}		& \footnotesize{[10, 0, 1996]} 	&\footnotesize{[8, 2, 1990]}		&\\
		&   	$42.51\pm 10.51$*				& $\bf 44.92\pm11.07 	$*		&	$ 10.63\pm3.15$				&$ 32.83\pm9.49$\\				
\tiny{DS5}& 	\footnotesize{[9, 0 ,1491]}		& \footnotesize{[10, 0, 1490]} 	&\footnotesize{[9, 1, 1490]}		&\\
		&   	 $73.75\pm 5.12$				& $\bf 78.76\pm 4.60	$*		&	$ 15.12\pm 13.05$				&$73.86\pm5.72$\\
\tiny{DS6}& 	\footnotesize{[9, 0 ,782]}		& \footnotesize{[10, 0, 781]} 	&	\footnotesize{[9, 1, 781]}		&\\
		&   	 $75.96\pm 4.89$				& $\bf 79.51\pm 5.33	$*		&	$ 38.86\pm 14.14$				&$76.03\pm4.77$\\
\end{tabular*}
\end{center}
\end{table}
\begin{table}\vspace{-1cm}
\begin{center}
\caption{\label{tab:comparison_50} Signature and average test set accuracy for SwissProt (DS1), Chromosome (DS2), Proteom (DS3), Zongker (DS4), Delft gestures (DS5), Woody (DS6) 
	using a Nystr\"om approximation with $10, 50, 100, full$ landmarks and no, clip or flip eigenvalue correction.}
\footnotesize
\begin{tabular*}{1\textwidth}{@{\extracolsep{\fill}}|c|c|c|c|c}\hline
		& 		$50$ / Clip				& 	$50$ / Flip       			& 	$50$ / No					& 	$50$ L-MDS   								\\\hline\hline
 \tiny{DS1}& 	 \footnotesize{[49, 0 ,10939]}	& \footnotesize{[50, 0, 10930]} 	&\footnotesize{[49, 1, 10931]}	&\\
		&  	 $76.21\pm 5.13$			& $76.49 \pm 3.73	$		&	$ 69.05\pm 5.01$					& $\bf 76.59 \pm 4.65$\\
 \tiny{DS2}&	\footnotesize{[49, 0 ,4151]}	& \footnotesize{[50, 0, 4150]} 	&\footnotesize{[49,1, 4150]}				&\\
		&   	 $94.05\pm1.17$			& $ 93.94\pm 1.28	$		&	$ 83.66 \pm 25.43$					&$\bf 94.11\pm 1.21$	\\
\tiny{DS3}& 	\footnotesize{[48, 0 ,2556]}	& \footnotesize{[50, 0, 2554]} 	&\footnotesize{[49, 1, 2550]}		&\\
		& 	 $93.08\pm2.25$			& $\bf 93.82\pm 1.59	$*		&	$ 3.53\pm3.25$					&$92.35\pm2.08$\\
\tiny{DS4}&	\footnotesize{[34, 0 ,1979]}	& \footnotesize{[50, 0, 1950]} 	&\footnotesize{[34, 16, 1950]}		&\\
		&   	 $80.79\pm3.94$*			& $\bf 85.35\pm 3.42$*			&	$9.82 \pm 2.08$				&$ 73.57\pm6.71$\\				
\tiny{DS5}& 	\footnotesize{[48,0,1452]}	& \footnotesize{[50, 0, 1450]} 	&\footnotesize{[48, 2, 1450]}	&\\
		&   	 $\bf 95.31\pm1.82$			& $ 94.72\pm 2.25	$		&	$24.99 \pm 27.56$				&	$ 95.31\pm1.89$\\
\tiny{DS6}& 	\footnotesize{[49, 0 ,742]}		& \footnotesize{[50, 0, 741]} 	&\footnotesize{[49, 1, 741]}		&\\
		&   	 $88.55\pm 4.11$				& $\bf 89.30\pm 3.72	$		&	$ 81.40\pm 23.63$				&$88.46\pm4.35$\\
\end{tabular*}
\end{center}
\end{table}
\begin{table}\vspace{-1cm}
\begin{center}
\caption{\label{tab:comparison_100} Signature and average test set accuracy for SwissProt (DS1), Chromosome (DS2), Proteom (DS3), Zongker (DS4), Delft gestures (DS5), Woody (DS6) 
	using a Nystr\"om approximation with $10, 50, 100, full$ landmarks and no, clip or flip eigenvalue correction. }
\footnotesize
\begin{tabular*}{1\textwidth}{@{\extracolsep{\fill}}|c|c|c|c|c}\hline
		& 		$100$ / Clip				& 	$100$ / Flip       			& 	$100$ / No   			& 	$100$ L-MDS					\\\hline\hline
 \tiny{DS1}& 	 \footnotesize{[99, 0 ,10889]}	& \footnotesize{[100, 0, 10888]} &\footnotesize{[99, 1, 10888]}	&\\
		&  	 $87.62\pm 2.11$			& $ 87.63\pm 1.85	$		&	$\bf 88.17 \pm 2.19$				&	$87.50 \pm 2.24$\\
 \tiny{DS2}&	\footnotesize{[91, 0 ,4109]}	& \footnotesize{[100, 0, 4100]} &\footnotesize{[91,9,4100]}			&\\
		&   	 $95.00\pm 1.11$			& $ 94.71\pm 1.68	$		&	$ 11.29\pm7.68$				&	$\bf 95.18\pm 1.07$	\\
\tiny{DS3}& 	\footnotesize{[96, 0 ,2506]}	& \footnotesize{[99, 0, 2505]}	 &\footnotesize{[97, 2, 2505]}	&	\\
		& 	 $96.48\pm 1.34$			& $\bf 96.96\pm 1.17	$		&	$ 13.75 \pm 9.90$				&	$96.29\pm1.27$	\\
\tiny{DS4}&	\footnotesize{[63, 0 ,1937]}	& \footnotesize{[100, 0, 1900]} &\footnotesize{[62, 38, 1900]}		&\\
		&   	 $83.47\pm4.31$*			& $\bf 87.42\pm 3.15	$*		&	$10.55 \pm 2.43$				&	$80.34\pm7.73$	\\				
\tiny{DS5}& 	\footnotesize{[91, 0 ,1401]}	& \footnotesize{[100, 0, 1400]} &\footnotesize{[92, 8, 1400]}	&\\
		&   	 $\bf 96.07\pm 1.56$			& $94.74\pm 4.23 	$		&	$ 23.33 \pm 18.62$				&	$ 96.01\pm1.69$\\
\tiny{DS6}& 	\footnotesize{[96, 0 ,695]}		& \footnotesize{[100, 0, 691]} 	&\footnotesize{[96, 4, 691]}		&\\
		&   	 $90.69\pm 3.38$				& $\bf 90.71\pm 3.20	$		&	$ 38.11\pm 23.74$				&$90.51\pm3.65$\\
\end{tabular*}
\end{center}
\end{table}

\begin{table*}[ht]
\begin{center}
\caption{\label{tab:comparison_full} Average test set accuracy for SwissProt (DS1), Chromosome (DS2), Proteom (DS3), Zongker (DS4), Delft gestures (DS5), Woody (DS6)
using the standard approach (no-approximations) and the flip, clip or no-eigenvalue correction on the full matrix. This has $\mathcal{O}(N^3)$ complexity.}
%\footnotesize
\begin{tabular*}{\textwidth}{@{\extracolsep{\fill}}l|c|c|c}\hline
Data set 	&	clip				& 	flip					&    no     \\\hline\hline
 DS1		&   	$ 95.45\pm 0.88$	& $ 95.39\pm 1.01$		& $ 95.40\pm0.59$  \\
 DS2		&    	$ 97.12\pm 0.89$	& $ 97.17\pm 0.99$		& $ 96.93\pm 0.66$  \\
 DS3		&     	$ 99.42\pm 0.66$	& $ 99.42\pm 0.45$		& $ 99.38\pm 0.61$  \\
 DS4		&     	$ 95.65\pm 1.13$	& $ 96.25\pm 0.75$		& $ 25.25\pm 4.78$  \\
 DS5		&     	$ 98.33\pm 1.67$	& $ 98.00\pm 0.94$		& $ 96.13\pm 1.43$  \\
 DS6		&     	$ 92.54\pm 2.27$	& $ 93.17\pm 2.48$		& $ 89.63\pm 3.58$  \\
\end{tabular*}
\end{center}
\end{table*}

\begin{table*}[ht]
\begin{center}
\caption{\label{tab:comparison_diss_space} Average test set accuracy for SwissProt (DS1), Chromosome (DS2), Proteom (DS3), Zongker (DS4), Delft gestures (DS5), Woody (DS6) 
using the dissimilarity space representation and a linear kernel or an elm kernel.}
%\footnotesize
\begin{tabular*}{\textwidth}{@{\extracolsep{\fill}}l|c|c}\hline
Data set 	&				linear		& 	elm        \\\hline\hline
 DS1		&  $26.01\pm5.49$  	& $72.09\pm 0.96$		\\
 DS2		&   $76.76\pm1.11$	& $89.88\pm 0.96$		\\
 DS3		&   $68.36\pm2.48$	& $85.37\pm 2.86$		\\
 DS4		&   $93.70\pm2.04$	& $95.05\pm 1.71$		\\
 DS5		&   $87.73\pm3.83$	& $91.67\pm 2.58$		\\
 DS6		&   $28.83\pm6.97$	& $89.38\pm 4.48$		\\
\end{tabular*}
\end{center}
\end{table*}		

\begin{figure*}
\caption{Spearman rank correlation (left) and the crossvalidation accuracy (right) 
for the three largest data sets using the proposed approach with an interleaved double centering 
and Nystr\"om approximation on the dissimilarity data. 
%Bottom a logarithmic representation 
%of the eigenspectrum of the unapproximated and double centered matrix.correl_pred
}\label{fig:correl_pred}

\begin{center}
	\subfigure[Swiss correlation]{\includegraphics[width=0.49\textwidth]{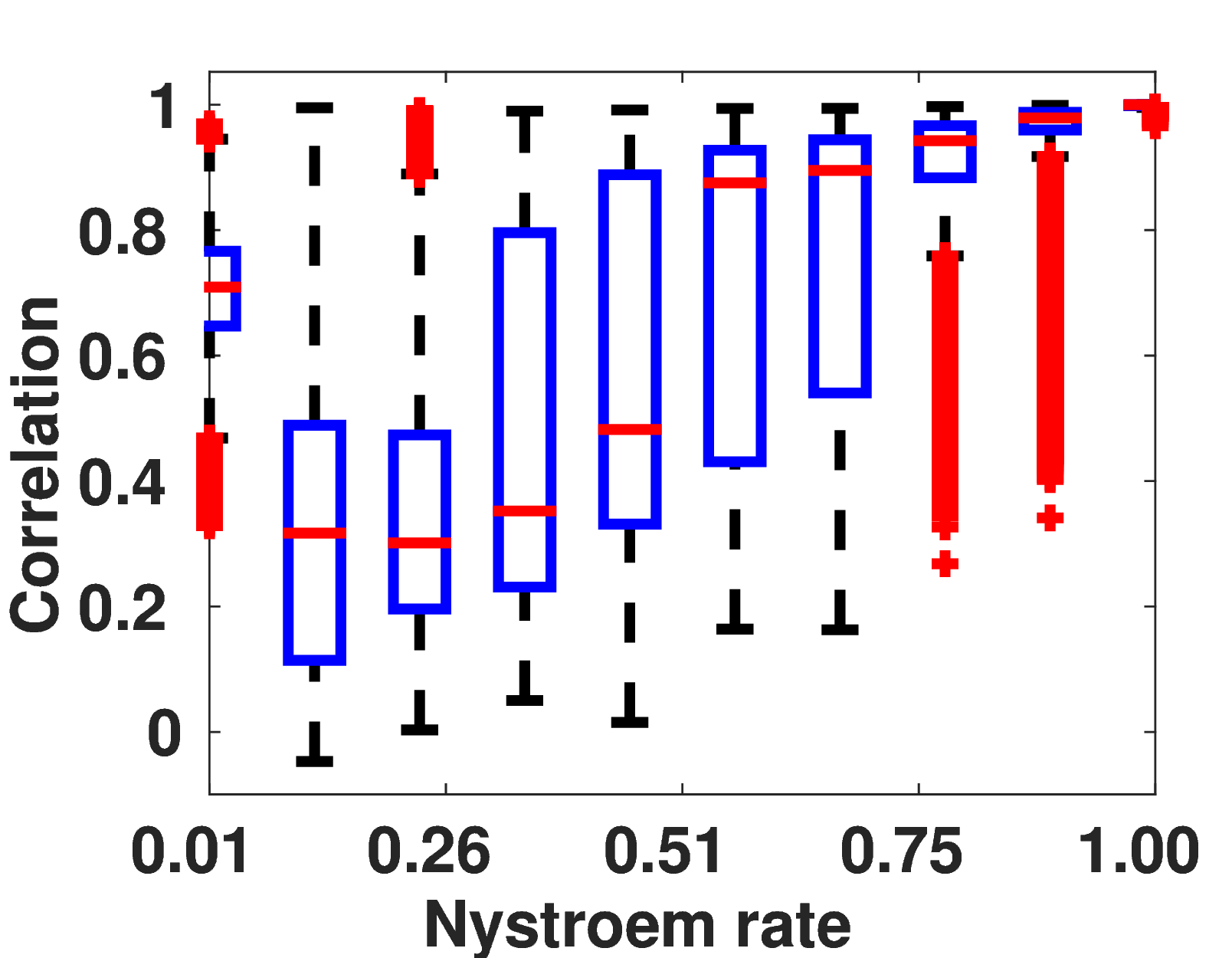}}
	\subfigure[Swiss accuracy]{\includegraphics[width=0.49\textwidth]{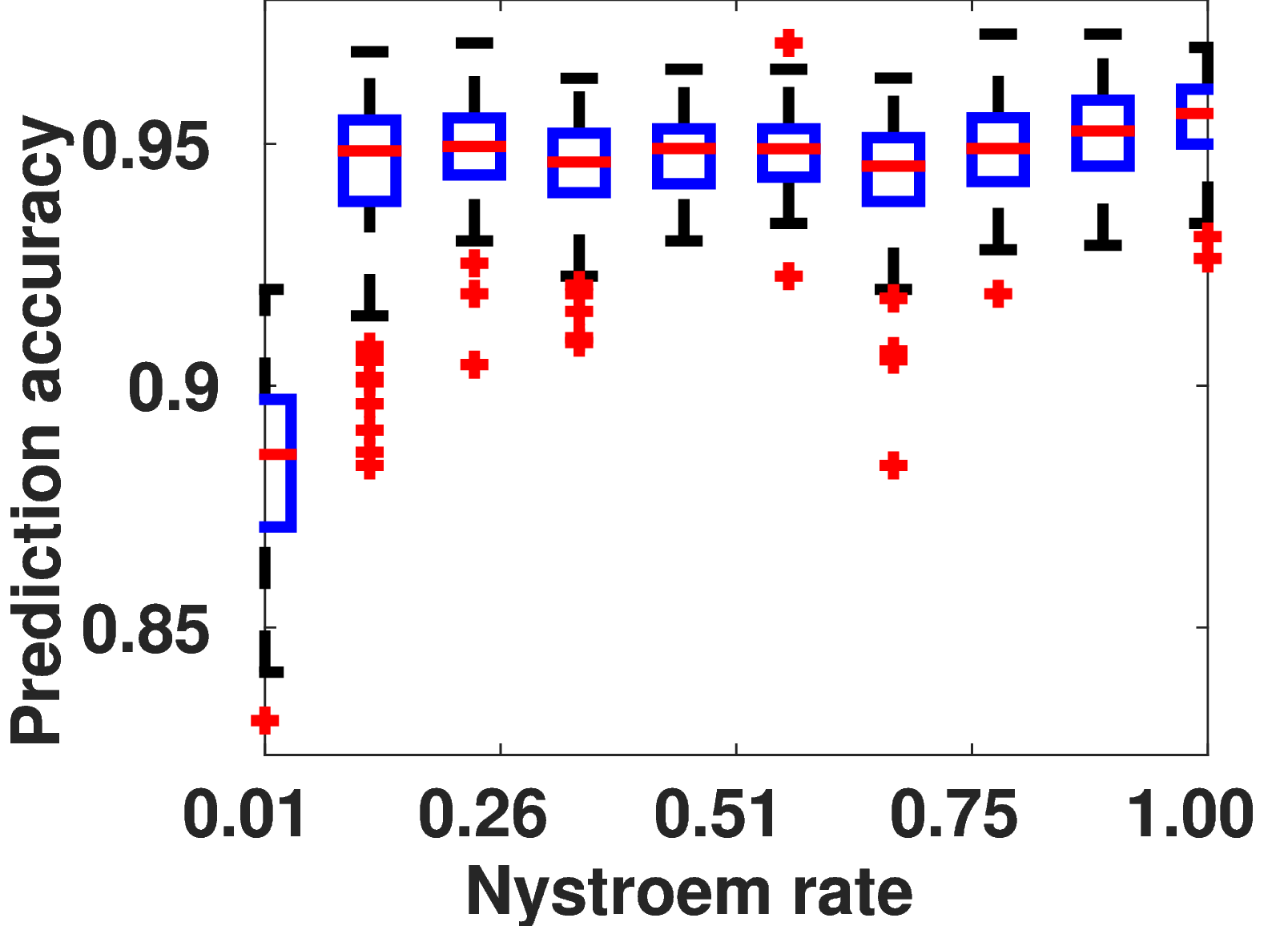}}\\
	\subfigure[Chromosom correlation]{\includegraphics[width=0.49\textwidth]{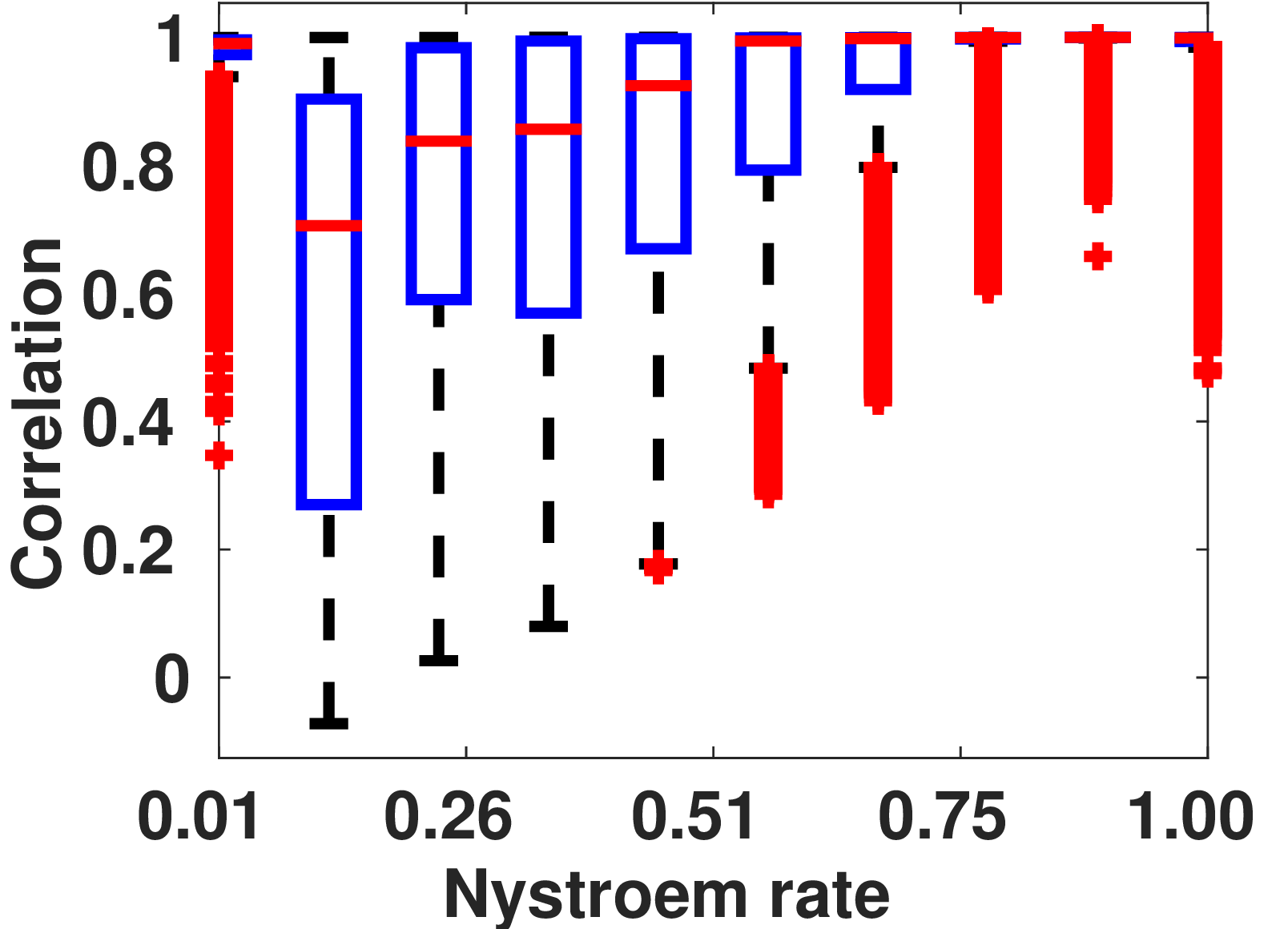}}
	\subfigure[Chromosom accuracy]{\includegraphics[width=0.49\textwidth]{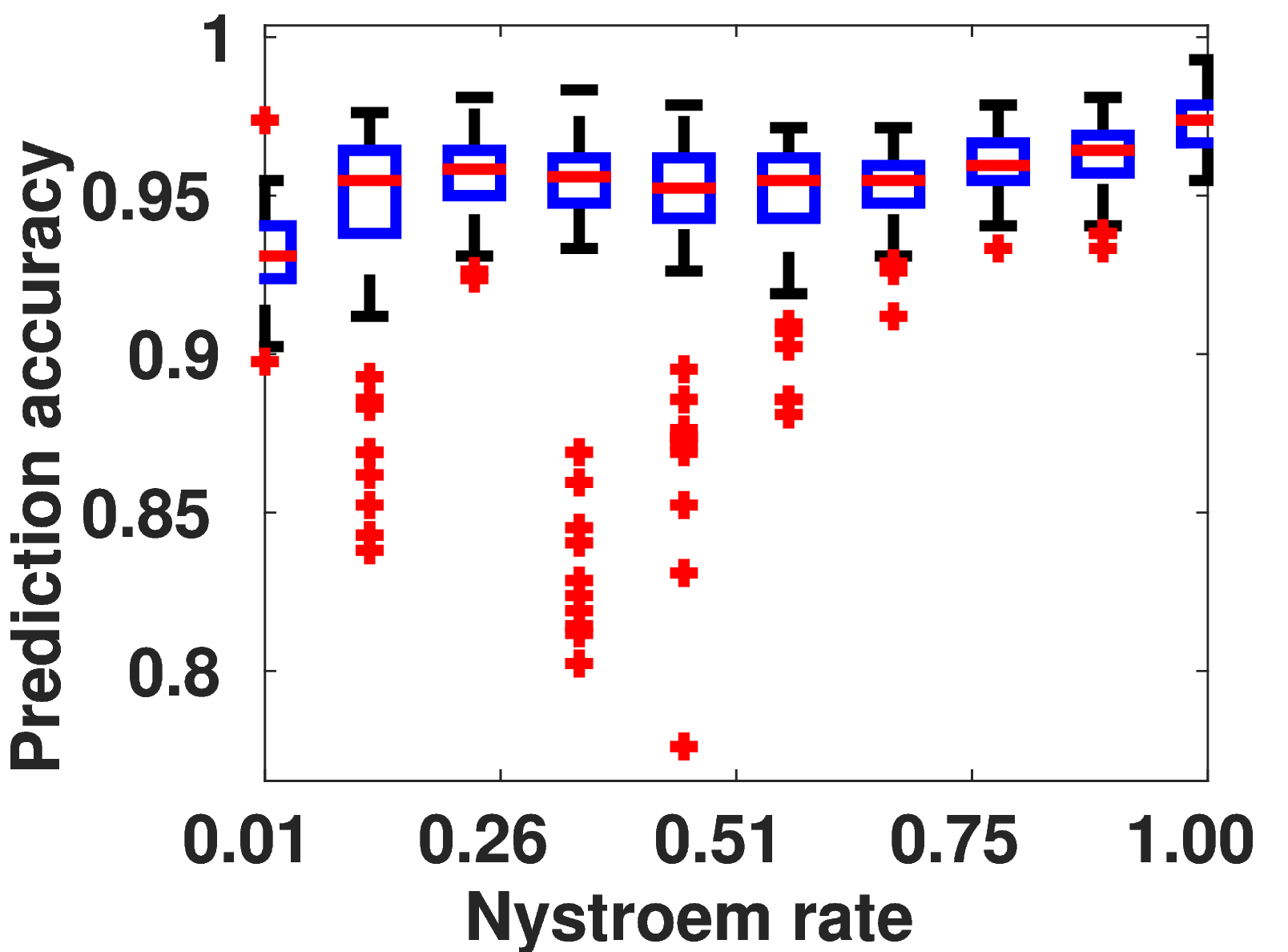}}\\
	\subfigure[Proteom correlation]{\includegraphics[width=0.49\textwidth]{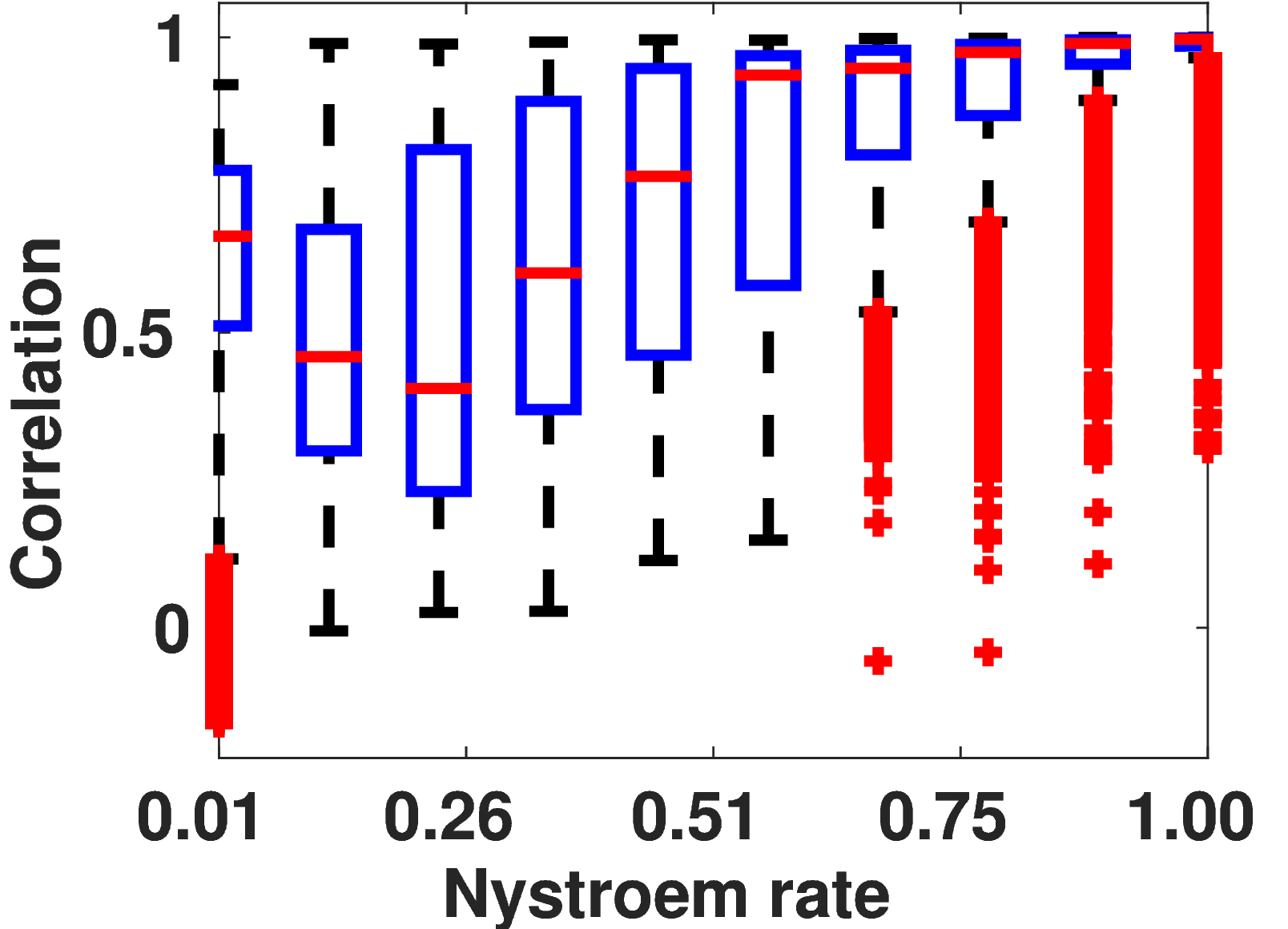}}
	\subfigure[Proteom accuracy]{\includegraphics[width=0.49\textwidth]{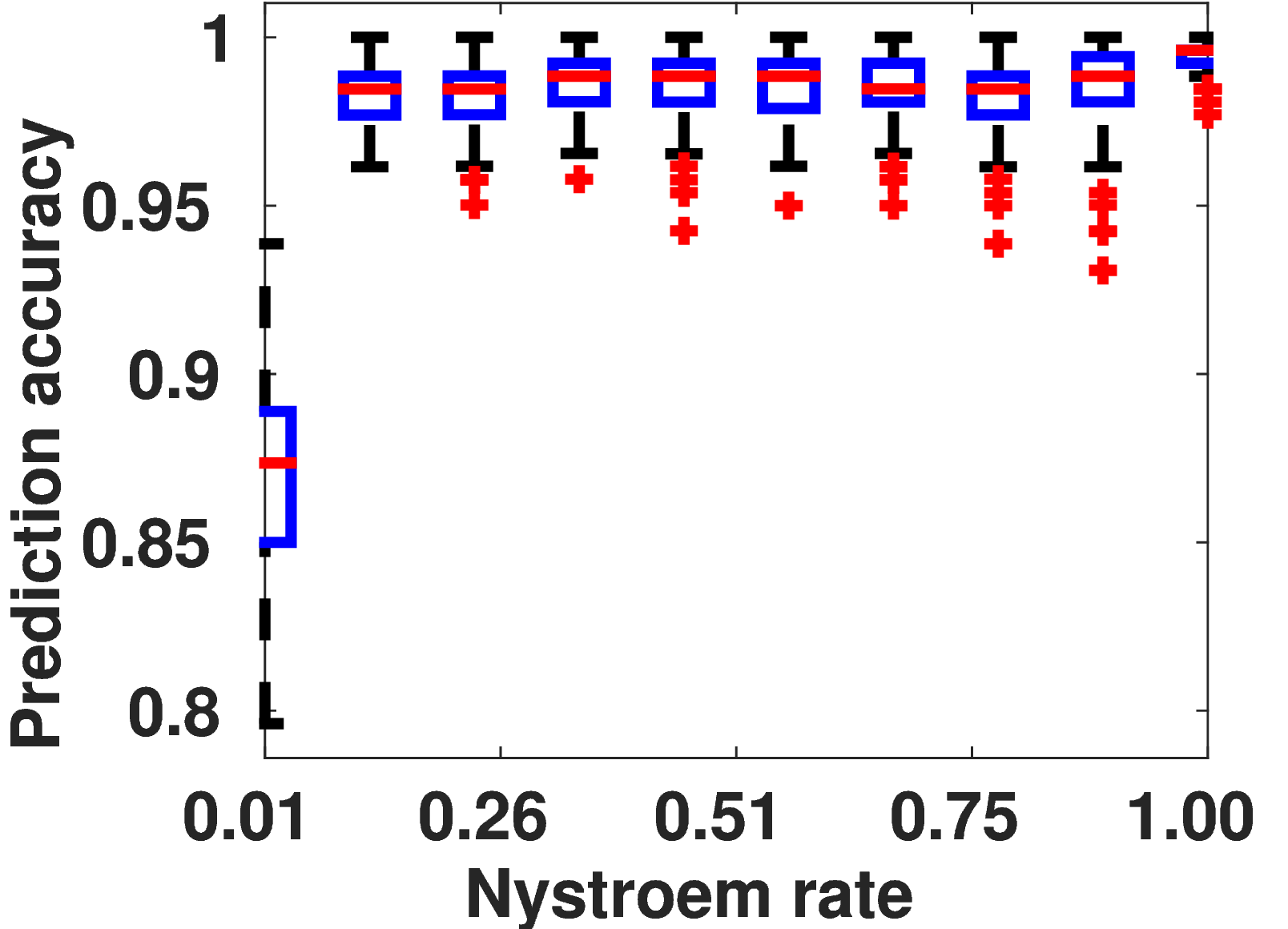}}
\end{center}
\end{figure*}

\begin{figure*}
\caption{Logarithmic representation of the eigenspectrum of the unapproximated and double centered matrix for the
larger datasets DS1 - DS3.}\label{fig:log_ev}
\begin{center}
	\subfigure[Swiss eigenspectrum]{\includegraphics[width=0.32\textwidth]{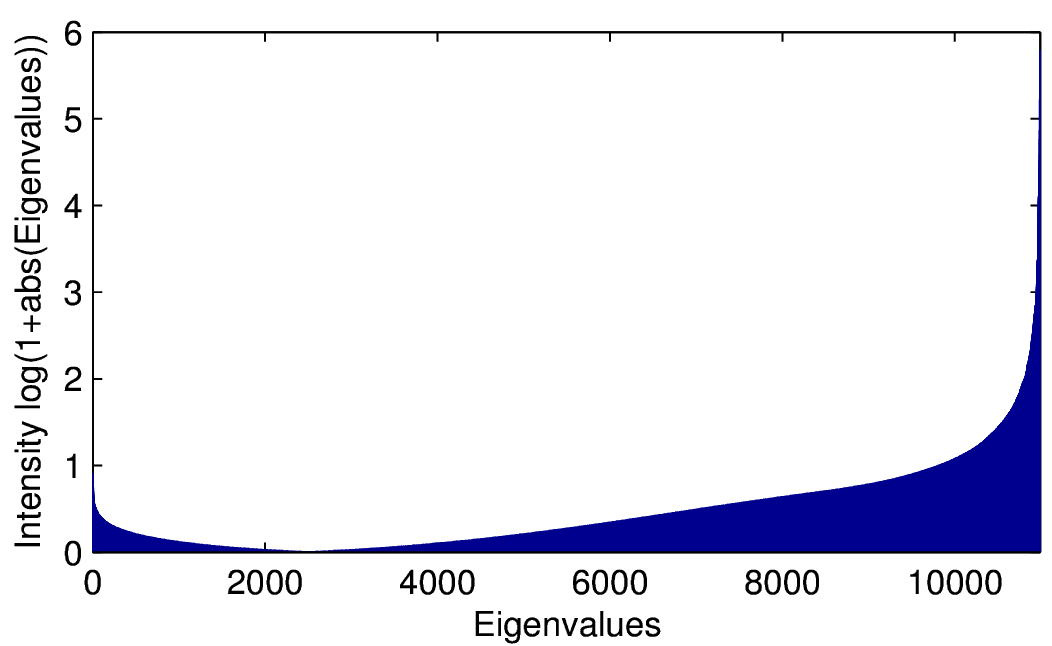}}
	\subfigure[Chromosom eigenspectrum]{\includegraphics[width=0.32\textwidth]{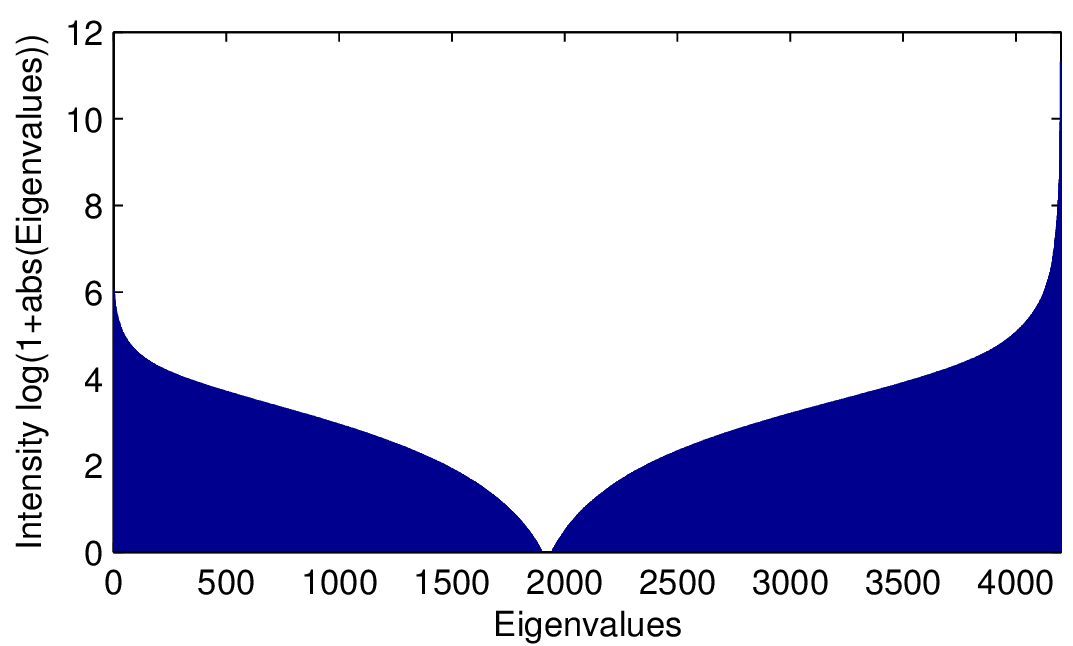}}
	\subfigure[Proteom eigenspectrum]{\includegraphics[width=0.32\textwidth]{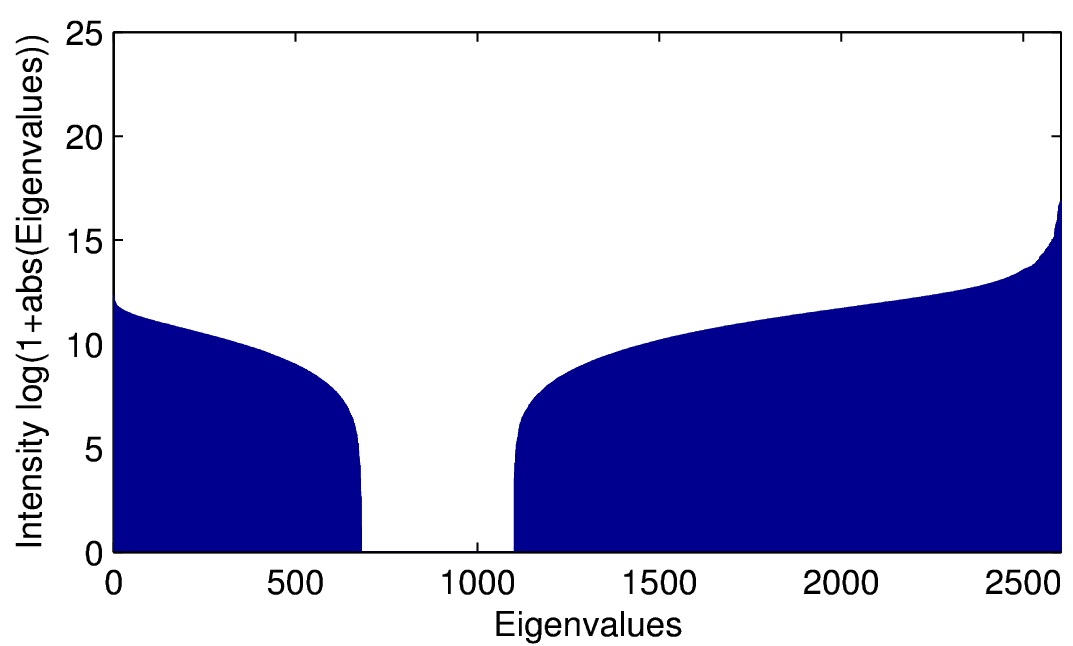}}
\end{center}
\end{figure*}

To get comparable experiments, the same randomly drawn landmarks are used in each of the corresponding sub-experiments
(along a column in the table). New landmarks are only drawn for different Nystr\"om approximations and for sample sizes shown in Figure \ref{fig:diss_rt_swiss}.
Classification rates are calculated in a 10-fold crossvalidation with 10 repeats using the Core-Vector-Machine (CVM) \cite{DBLP:conf/icml/TsangKK07}. 
The crossvalidation does not include a new draw of the landmarks, to cancel out the selection bias of the Nystr\"om approximation, accordingly CVM 
use the same kernel matrices. However, our objective is not maximum classification performance (which is only one possible
application) but to demonstrate the effectiveness of our approach for dissimilarity data of larger scale. 

First, one observes that the eigenvalue correction has a strong, positive effect
on the classification performance consistent with earlier findings \cite{DBLP:journals/jmlr/ChenGGRC09}. 
Best results over a row are highlighted in bold at the various result tables. If the difference is significantly better than L-MDS
a $\star$ has been added. 
Raising the number of landmarks improves the classification performance for the experiments with eigenvalue correction. 
Using kernels without eigenvalue correction has in general a negative impact. While an increase in the number of landmarks leads to a better
approximation of the dataset and may therefore improve the classification accuracy it can also raise the influence of negative eigenvalues,
damping the performance\footnote{Comparing signatures at different Nystr\"om approximations also shows that many 
eigenvalues are close to zero and are sometimes counted as positive,negative or zero.}. We  found that flipping is in general
superior to clipping. For $m=10$ flipping was consistently better than clipping or L-MDS. With an increase of $m$ the approximation error
of L-MDS vanishes and the results become more and more similar to the clipping results. But for DS4 L-MDS is also inferior if $m=100$,
which shows that for some data L-MDS gives bad results, due to its approximation errors even for rather large $m$.
Especially for DS3,DS4 and DS6 we observe that the proposed method gives much better results.

In Table \ref{tab:comparison_diss_space} we also show the
crossvalidation results by use of the priorly mentioned dissimilarity space representation. For simplicity we use an $N$ dimensional feature space 
and analyse the obtained vector representation by means of a linear kernel and a defacto parameter free elm kernel as proposed by \cite{DBLP:journals/ijon/FrenayV11}.
For the majority of the experiments the obtained results are significantly worse with the exception of DS4. Also for DS5 a comparison with the Nystr\"om
approximation at $m=100$ gives still acceptable results. It should be noted that the results of the elm-kernel experiments are consistently better compared to
the linear kernel, indicating the high non-linearity of the data. Obviously the dissimilarity space representation is in general no reasonable alternative.
Additionally it becomes very costly for out-of-sample extensions if the number of considered features is large.

In another experiment, see Figure \ref{fig:correl_pred} we analyzed the proximity preservation of the approximated and corrected
matrix with respect to the unapproximated and corrected matrix. One would expect that for very low Nystr\"om rates (high approximation),
only the dominating eigenvalues are kept and the approximation suffers mainly when the eigenspectra are very smooth. At
increasing Nystr\"om rates (lower approximation), first more and more small eigenvalues (also negative ones) are kept leading to 
a more complex data set and accordingly also a more complex proximity preservation task. Finally if the Nystr\"om rates are high
(almost no approximation) one would expect a perfect preservation. This effect is indeed observed in Figure \ref{fig:correl_pred}.
We used the Spearman's rank correlation to measure how far the ranks of the proximities (e.g. distances) are preserved between
the two approaches, namely our proposal and a full double centering, followed by a full eigenvalue correction.
Low correlation indicates that the data relations are not well preserved whereas small correlation errors 
indicate that most likely only local neighborhood relations are confused. Comparing the correlation results (left plots in Figure \ref{fig:correl_pred})
with the prediction accuracy on the test data (right plots in Figure \ref{fig:correl_pred}) we see that only strong variations in the
correlation lead to strong misclassifications. This agrees with our expectation that the data are potentially clustered and local
errors in the data relation have only a weak or no effect on the classification model. Similar results were found if we compare our approach
to data which have been first double-centered without approximations and where only the eigenvalue correction is done using the Nystr\"om
approach.

From the analysis we can conclude that the proposed approach is quite effective to keep the global relations in the data space
also for quite high approximations, which is relevant for classification and clustering the data. The local neighborhood relations 
are kept only for approximation rates of above $60\%$. As one can see from smooth eigenspectra in Figure \ref{fig:log_ev}, the 
rank of the data sets is rather high, accordingly only for large $m$ the approximation can keep detail information, effecting the
local relationships of the data points. Thus, if the different classes are close to each other
and have complex nonlinear boundaries,
decreasing the number of landmarks
leads to an increased classification error.
In practice, as can be seen on the Figure \ref{fig:correl_pred},
the number of the landmarks needs to be very small to take effect.
It is thus possible to approximate the matrices
by selecting $m$ sufficiently small,
without sacrificing the classification accuracy.

\subsection{Runtime performance}
As shown exemplary in Figure \ref{fig:diss_rt_swiss} the classification performance on eigenvalue-cor\-rected data is approximately 
the same for our proposed strategy and the standard approach.
But the runtime performance is  drastically better for an 
increase in the number of samples.
To show this we selected subsets from the considered data with different sizes
from $1000$ to the maximal number, while the number of landmarks is fixed by $L=500$ and calculated the runtime and classification
performance using the CVM classifier in a 10-fold crossvalidation. The eigenvalues have been flipped in this experiment.
The results of the proposed approach compared to the standard approach are shown in the plots of Figure \ref{fig:diss_rt}. 
For larger $N$ the runtime of the standard method (red/dashed line) is two magnitudes larger on log-scale compared to the proposed approach.

\section{Large scale experiments}
As a final experiment we analyze the proposed approach for large scale non-metric proximity data. With respect
to the work presented in the former sections a valid application of kernel methods for such data is not yet possible.
Neither the classical eigenvalue correction approach  \cite{DBLP:journals/jmlr/ChenGGRC09} nor the learning
of a proximity kernel \cite{DBLP:conf/icml/ChenGR09} scales to larger data sets with $N \gg 1e3$ samples,
the problem becomes even more challenging if the data are given as dissimilarities such that a double
centering is needed to keep a corresponding representation. Due to the large number of samples a full matrix
reconstruction is not any longer possible to calculate error measures like the Spearman rank correlation accordingly
we only provide test set errors obtained within a $10$ fold crossvalidation using a CVM.
In our experiments we consider:
\begin{itemize}
	\item The SwissProt protein database \cite{swissprot} but now at \emph{larger scale} in the version of 11/2010,
restricted to ProSite labeled sequences with at least $1,000$ entries per label. We obtain 
$46$ ProSite labels and $82,525$ sequences which are compared by the Smith-Waterman alignment
algorithm as provided in \cite{citeulike:668527}. We refer to this data as DS-L-1. The obtained similarity scores 
are symmetric but non-metric,  accordingly standard kernel methods can not be used directly in a valid form.
We take $1,000$ landmarks, randomly taken from the selected classes. 
The dataset has $2$ larger negative eigenvalues in the approximated matrix.
	\item The Pavia remote sensing data consist of $42.776$ spectra (DS-L-2). The dataset is taken from \cite{RemoteSensing}. 
	We use the symmetrized Kullback-Leibler Divergence, which is also known as the 
	spectral information divergence (SID) in remote sensing and frequently used as an effective \emph{non-metric} 
	measure to compare spectral data \cite{vanderMeer20063} and use $10\%$ randomly chosen points as landmarks.

	\item The Salina data of $54129 $ points (DS-L-3) also taken from \cite{RemoteSensing} with the same measure
	and settings as for DS-L-2
	\item The ball dataset with $30,000$ samples (Ball-Large). Landmarks are selected randomly as $10\%$ from the dataset.
\end{itemize}
For all of these data sets a standard kernel approach is costly in calculating the whole similarity matrix 
and it would be basically impossible to get an eigenvalue correction in a reasonable time.
Modern kernel classifiers like the Core-Vector Machine (CVM)\cite{DBLP:conf/icml/TsangKK07}
do not need to evaluate all the kernel similarities but our similarities are non-metric and an 
accurate online eigenvalue correction is not available. 
However we can use our presented approach approximating the score matrix as well as performing
an eigenvalue correction. The calculation of the final approximated kernel function  and eigenvalue correction by the presented approach 
takes only some minutes. 

The obtained approximated and now positive semi definite similarity
matrices can  be used by a Core-Vector Machine in a $10$ fold crossvalidation to generate a 
classification model with a good mean prediction accuracy see Table \ref{tab:results_large}.
\begin{table}\centering
\begin{tabular*}{\textwidth}{@{\extracolsep{\fill}}l|c|c|c|c|c|c}
	Data		& 	size 	 & type	&	flip						& clip				& No				&	L-MDS (clip)	\\\hline
	DS-L-1	&	80k	& S		& 	$\bf 96.24 \pm 0.29 \%$			& $96.22 \pm 0.28\%$	& \text{failed}  			&	$96.14\pm 0.27\%$\\
	DS-L-2	&	40k	& D		& 	$\bf 82.56 \pm 0.60 \%$			& $79.80 \pm 0.94\%$	& \text{failed}  			&	$81.18\pm 1.17\%$		\\	
	DS-L-3	&	50k	& D		& 	$\bf 88.11 \pm 0.68\%$*			& $85.06 \pm 0.73\%$	& \text{failed}  			&	$81.37\pm 0.62\%$	\\
	Ball-Large	&      30k	& D 		&	$\bf 93.59 \pm  0.63\%$*			& $50.28 \pm 0.80\%$	& $28.50 \pm 0.76\%$ 	&	$50.13 \pm 0.97\%$\\	
\end{tabular*}
	\caption{Crossvalidation results of the large scale data sets (D - dissimilarities, S - similarities) using flip, clip or no eigenvalue correction\label{tab:results_large}.}
\end{table}
An additional benefit of the CVM approach is that it naturally leads to very sparse models. Accordingly
the out of sample extension to new sequences requires only few score calculations to the sequences 
of the training set.
%
%Using the same encoded similarity matrix and the approach used in \cite{DBLP:conf/icml/LiKL10} 
%we can now, having a \emph{psd} similarity matrix, calculate a laplacian eigenmap \cite{DBLP:conf/nips/BelkinN01} of our large scale datasets. 
%Exemplary we show the result for DS-L-1 in Figure \ref{fig:swissprot_zoom_21}.
%\begin{figure}
%	\centering
%	\includegraphics[width=0.6\columnwidth]{swissprot_sw_largest_1000_flip_centered_zoom}
%	\caption{Eigenmap of the DS-L-1 data using the flip eigenvalue correction.}\label{fig:swissprot_zoom_21}\vspace{-0.5cm}
%\end{figure}
%This visualization can be used to analyze local relations of the different sequences.
%
\section{Conclusions}
In this article we addressed the analysis of potentially non-metric proximity data and especially the relation between dissimilarity and similarity data.
We proposed effective and \emph{accurate} transformations across the different representations. The results show that our approach can be
understood as a generalization of Landmark MDS. L-MDS did not show any significant superior results compared to our method, but instead was often
found to be significantly worse. This finding also persisted if the number of landmarks was raised to a rather large value.
 
Dedicated learning algorithms for dissimilarities and kernels are now accessible for both types of data. 
The specific coupling of double centering and Nystr\"om approximation permits to compute
an exact eigenvalue decomposition in linear time which is a valuable result for many different methods
depending on the exact calculation of eigenvalues and eigenvectors of a proximity matrix. 
While our strategy is very effective e.g. to improve supervised learning of non-metric dissimilarities by kernel methods,
it is however also limited again by the Nystr\"om approximation, which itself may fail to provide sufficient approximation
and accordingly further research in this line is of interest. Nevertheless, dedicated methods for arbitrary proximity data 
as addressed in  \cite{DBLP:conf/sspr/PekalskaDGB04} will also be subject of future work. For non-psd data the error introduced by the Nystr\"om approximation and the eigenvalue correction 
is not yet fully understood and bounds similar as proposed in \cite{DBLP:journals/jmlr/DrineasM05}
are still an open issue. It is also of interest to extend our approach to other types of matrix approximation schemes
as e.g. the CUR algorithm and others \cite{wang2013improving,DBLP:conf/aistats/WangZ14,DBLP:conf/kdd/WangZQZ14}. In future work we will also analyze in more detail the handling of extremely large (dis-)similarity sets 
\cite{Schleif2014e,DBLP:conf/icml/GittensM13} and analyze our approach in the context of unsupervised problems \cite{DBLP:conf/icml/ZhangTK08}.

\section*{Acknowledgments}
We would like to thank Alexander Grigor'yan,  Faculty of Mathematics, 
University of Bielefeld for effective discussions about functional analysis and eigenspaces
and Barbara Hammer, Center of Excellence, University of Bielefeld for continuous support in this project.
%Both authors are grateful to the \emph{Max-Planck-Institute for Physics of Complex Systems in Dresden} 
%and Michael Biehl, Thomas Villmann and Manfred Opper as the organizer of the 
%\emph{Statistical Inference: Models in Physics and Learning}-Workshop for providing a nice working atmosphere during
%the preparation of this manuscript.
Financial support from the Cluster of Excellence 277 Cognitive Interaction Technology 
funded by the German Excellence Initiative is gratefully acknowledged.  F.-M. Schleif was supported by a Marie Curie Intra-European Fellowship (IEF): FP7-PEOPLE-2012-IEF (FP7-327791-ProMoS) 
We would like to thank R. Duin and E. Pekalska for providing access to some of the non-metric datasets and the distools and prtools.

\section{Appendix}
\textbf{Definition:}
The norm of an operator $K:L^2(\Omega) \to L^2(\Omega)$ is defined as
\[
\|K\|_{L^2 \to L^2}=\sup_{\|f\|\leq 1} \|K f\|_{L^2}
\]
and the norm of a function $f\in L^2(\Omega)$ is defined as
\[
\|f\|_{L^2} = \left(\int_\Omega |f(x)|^2 d\mu(x)\right)^{1/2}.
\]

\textbf{Theorem:}
The sequence of operators $K_m$ converges uniformly to $K$
in the operator norm if
\[
\sup_{\substack{x \in \Omega \\ y \in \Omega}}\left| k_m(x,y) - k(x,y) \right|
\leq \delta_m
\]
and $\delta_m \to 0$ for $m \to \infty$.

\textbf{Proof:}
The uniform convergence is given if
$\|K_m-K\|_{L^2 \to L^2} \to 0$
for $m \to \infty$.
Thus, we need to compute this quantity.
Following the computations in \cite{werner},
we can write for the norm of $Kf$
\begin{align*}
\|Kf\|^2_{L^2}
= & \int_\Omega |K f(x)|^2 d\mu(x)\\
= & \int_\Omega \left|\int_\Omega k(x,y) f(y) d\mu(y)\right|^2 d\mu(x)\\
\leq & \int_\Omega \left(\int_\Omega |k(x,y)| |f(y)| d\mu(y)\right)^2 d\mu(x)\\
\leq & \int_\Omega \left(\int_\Omega |k(x,y)|^2  d\mu(y)\right)
  \left(\int_\Omega |f(y)|^2 d\mu(y)\right) d\mu(x)\\
= & \int_\Omega \int_\Omega |k(x,y)|^2 d\mu(x) d\mu(y) \|f\|^2_{L^2}
\end{align*}
where we used H\"older's inequality and Fubini's theorem.
It follows
\begin{align*}
\|K_m-K\|_{L^2 \to L^2}^2
= & \sup_{\|f\|\leq 1} \|(K_m - K) f\|_{L^2}^2 \\
= & \sup_{\|f\|\leq 1} \int_\Omega |(K_m - K) f(x)|^2 d\mu(x)\\
\leq & \sup_{\|f\|\leq 1} 
  \int_\Omega \int_\Omega |k_m(x,y) - k(x,y)|^2 d\mu(x) d\mu(y) \|f\|^2_{L^2} \\
\leq & \int_\Omega \int_\Omega \delta_m^2 d\mu(x) d\mu(y) \\
= & \delta_m^2
\end{align*}
and since $\delta_m \to 0$ for $m \to \infty$,
we have $\|K_m-K\|_{L^2 \to L^2} \to 0$
for $m \to \infty$.
$\Box$

\bibliographystyle{plain}
\bibliography{ieee_tnnls_bib,simbad,pub_fms}

\begin{thebibliography}{10}

\bibitem{DBLP:journals/comgeo/BadoiuC08}
Mihai Badoiu and Kenneth~L. Clarkson.
\newblock Optimal core-sets for balls.
\newblock {\em Comput. Geom.}, 40(1):14--22, 2008.

\bibitem{DBLP:journals/ml/BalcanBS08}
Maria-Florina Balcan, Avrim Blum, and Nathan Srebro.
\newblock A theory of learning with similarity functions.
\newblock {\em Machine Learning}, 72(1-2):89--112, 2008.

\bibitem{DBLP:conf/eccv/BelongieFCM02}
Serge Belongie, Charless Fowlkes, Fan R.~K. Chung, and Jitendra Malik.
\newblock Spectral partitioning with indefinite kernels using the {N}ystr{\"o}m
  extension.
\newblock In Anders Heyden, Gunnar Sparr, Mads Nielsen, and Peter Johansen,
  editors, {\em ECCV (3)}, volume 2352 of {\em Lecture Notes in Computer
  Science}, pages 531--542. Springer, 2002.

\bibitem{Berg1984}
Christian Berg, Jens Peter~Reus Christensen, and Paul Ressel.
\newblock {\em Harmonic Analysis on Semigroups}.
\newblock Springer-Verlag, 1984.

\bibitem{swissprot}
B.~Boeckmann, A.~Bairoch, R.~Apweiler, M.-C. Blatter, A.~Estreicher,
  E.~Gasteiger, M.J. Martin, K.~Michoud, C.~O'Donovan, I.~Phan, S.~Pilbout, and
  M.~Schneider.
\newblock The swiss-prot protein knowledgebase and its supplement trembl in
  2003,.
\newblock {\em Nucleic Acids Research}, 31:365--370.

\bibitem{DBLP:journals/siammax/BrickellDST08}
Justin Brickell, Inderjit~S. Dhillon, Suvrit Sra, and Joel~A. Tropp.
\newblock The metric nearness problem.
\newblock {\em {SIAM} J. Matrix Analysis Applications}, 30(1):375--396, 2008.

\bibitem{DBLP:journals/jmlr/ChenGGRC09}
Yihua Chen, Eric~K. Garcia, Maya~R. Gupta, Ali Rahimi, and Luca Cazzanti.
\newblock Similarity-based classification: Concepts and algorithms.
\newblock {\em JMLR}, 10:747--776, 2009.

\bibitem{DBLP:conf/icml/ChenGR09}
Yihua Chen, Maya~R. Gupta, and Benjamin Recht.
\newblock Learning kernels from indefinite similarities.
\newblock In {\em In Proceedings of the 26th Annual International Conference on
  Machine Learning, ICML 2009, Montreal, Quebec, Canada, June 14-18, 2009},
  page~19, 2009.

\bibitem{DBLP:conf/kdd/ChittaJHJ11}
Radha Chitta, Rong Jin, Timothy~C. Havens, and Anil~K. Jain.
\newblock Approximate kernel k-means: solution to large scale kernel
  clustering.
\newblock In {\em Proceedings of the 17th ACM SIGKDD International Conference
  on Knowledge Discovery and Data Mining, San Diego, CA, USA, August 21-24,
  2011}, pages 895--903, 2011.

\bibitem{Cichocki20101532}
A.~Cichocki and S.-I. Amari.
\newblock Families of alpha- beta- and gamma- divergences: Flexible and robust
  measures of similarities.
\newblock {\em Entropy}, 12(6):1532--1568, 2010.

\bibitem{DBLP:journals/jmlr/CortesMT10}
Corinna Cortes, Mehryar Mohri, and Ameet Talwalkar.
\newblock On the impact of kernel approximation on learning accuracy.
\newblock {\em JMLR - Proceedings Track}, 9:113--120, 2010.

\bibitem{DBLP:conf/nips/SilvaT02}
Vin de~Silva and Joshua~B. Tenenbaum.
\newblock Global versus local methods in nonlinear dimensionality reduction.
\newblock In {\em Advances in Neural Information Processing Systems 15 [Neural
  Information Processing Systems, NIPS 2002, December 9-14, 2002, Vancouver,
  British Columbia, Canada]}, pages 705--712, 2002.

\bibitem{DBLP:journals/jmlr/DrineasM05}
Petros Drineas and Michael~W. Mahoney.
\newblock On the nystr{\"o}m method for approximating a gram matrix for
  improved kernel-based learning.
\newblock {\em Journal of Machine Learning Research}, 6:2153--2175, 2005.

\bibitem{PrTools:2012:Online}
R.~P.W. Duin.
\newblock {PRT}ools, march 2012.

\bibitem{DBLP:conf/sspr/DuinP10}
Robert P.~W. Duin and Elzbieta Pekalska.
\newblock Non-euclidean dissimilarities: Causes and informativeness.
\newblock In {\em Structural, Syntactic, and Statistical Pattern Recognition,
  Joint IAPR International Workshop, SSPR{\&}SPR 2010, Cesme, Izmir, Turkey,
  August 18-20, 2010. Proceedings}, pages 324--333, 2010.

\bibitem{DBLP:journals/jmlr/FarahatGK11}
Ahmed~K. Farahat, Ali Ghodsi, and Mohamed~S. Kamel.
\newblock A novel greedy algorithm for nystr{\"o}m approximation.
\newblock {\em JMLR - Proceedings Track}, 15:269--277, 2011.

\bibitem{DBLP:journals/pami/FowlkesBCM04}
Charless Fowlkes, Serge Belongie, Fan R.~K. Chung, and Jitendra Malik.
\newblock Spectral grouping using the {N}ystr{\"o}m method.
\newblock {\em IEEE Trans. Pattern Anal. Mach. Intell.}, 26(2):214--225, 2004.

\bibitem{DBLP:journals/ijon/FrenayV11}
Beno\^{\i}t Fr{\'e}nay and Michel Verleysen.
\newblock Parameter-insensitive kernel in extreme learning for non-linear
  support vector regression.
\newblock {\em Neurocomputing}, 74(16):2526--2531, 2011.

\bibitem{RemoteSensing}
Computational Intelligence~Group from~the Basque~University.
\newblock Hyperspectral remote sensing scenes, june 2014.

\bibitem{nips10gismokham}
A.~Gisbrecht, B.~Mokbel, and B.~Hammer.
\newblock The {N}ystrom approximation for relational generative topographic
  mappings.
\newblock In {\em NIPS Workshop}, 2010.

\bibitem{Schleif2012k}
A.~Gisbrecht, B.~Mokbel, F.-M. Schleif, X.~Zhu, and B.~Hammer.
\newblock Linear time relational prototype based learning.
\newblock {\em Journal of Neural Systems}, 22(5):online, 2012.

\bibitem{DBLP:journals/corr/abs-1303-1849}
Alex Gittens and Michael~W. Mahoney.
\newblock Revisiting the nystrom method for improved large-scale machine
  learning.
\newblock {\em CoRR}, abs/1303.1849, 2013.

\bibitem{DBLP:conf/icml/GittensM13}
Alex Gittens and Michael~W. Mahoney.
\newblock Revisiting the nystrom method for improved large-scale machine
  learning.
\newblock In {\em ICML (3)}, volume~28 of {\em JMLR Proceedings}, pages
  567--575. JMLR.org, 2013.

\bibitem{Goldfarb1984575}
L.~Goldfarb.
\newblock A unified approach to pattern recognition.
\newblock {\em Pattern Recognition}, 17(5):575 -- 582, 1984.

\bibitem{DBLP:journals/neco/GraepelO99}
Thore Graepel and Klaus Obermayer.
\newblock A stochastic self-organizing map for proximity data.
\newblock {\em Neural Computation}, 11(1):139--155, 1999.

\bibitem{Haasdonk2005482}
B.~Haasdonk.
\newblock Feature space interpretation of svms with indefinite kernels.
\newblock {\em IEEE Transactions on Pattern Analysis and Machine Intelligence},
  27(4):482--492, 2005.

\bibitem{DBLP:journals/neco/HammerH10}
Barbara Hammer and Alexander Hasenfuss.
\newblock Topographic mapping of large dissimilarity data sets.
\newblock {\em Neural Computation}, 22(9):2229--2284, 2010.

\bibitem{Jain19971386}
A.K. Jain and D.~Zongker.
\newblock Representation and recognition of handwritten digits using deformable
  templates.
\newblock {\em IEEE Transactions on Pattern Analysis and Machine Intelligence},
  19(12):1386--1391, 1997.

\bibitem{DBLP:conf/nips/KarJ11}
Purushottam Kar and Prateek Jain.
\newblock Similarity-based learning via data driven embeddings.
\newblock In {\em Proc. of Advances in Neural Information Processing Systems
  24: 25th Annual Conference on Neural Information Processing Systems 2011,
  Granada, Spain}, pages 1998--2006, 2011.

\bibitem{DBLP:conf/nips/KarJ12}
Purushottam Kar and Prateek Jain.
\newblock Supervised learning with similarity functions.
\newblock In {\em Proc. of Advances in Neural Information Processing Systems
  25: 26th Annual Conference on Neural Information Processing Systems 2012,
  Lake Tahoe, Nevada, United States}, pages 215--223, 2012.

\bibitem{mediansom}
T.~Kohonen and P.~Somervuo.
\newblock How to make large self-organizing maps for nonvectorial data.
\newblock {\em Neural Networks}, 15(8-9):945--952, 2002.

\bibitem{DBLP:journals/jmlr/KumarMT12}
Sanjiv Kumar, Mehryar Mohri, and Ameet Talwalkar.
\newblock Sampling methods for the nystr{\"o}m method.
\newblock {\em Journal of Machine Learning Research}, 13:981--1006, 2012.

\bibitem{DBLP:phd/de/Laub2004}
Julian Laub.
\newblock {\em Non-metric pairwise proximity data}.
\newblock PhD thesis, 2004.

\bibitem{DBLP:journals/pr/LaubRBM06}
Julian Laub, Volker Roth, Joachim~M. Buhmann, and Klaus-Robert M{\"u}ller.
\newblock On the information and representation of non-euclidean pairwise data.
\newblock {\em Pattern Recognition}, 39(10):1815--1826, 2006.

\bibitem{DBLP:conf/icml/LiKL10}
Mu~Li, James~T. Kwok, and Bao-Liang Lu.
\newblock Making large-scale nystr{\"o}m approximation possible.
\newblock In {\em Proceedings of the 27th International Conference on Machine
  Learning (ICML-10), June 21-24, 2010, Haifa, Israel}, pages 631--638, 2010.

\bibitem{DBLP:journals/jmlr/LiZY09}
Wu-Jun Li, Zhihua Zhang, and Dit-Yan Yeung.
\newblock Latent wishart processes for relational kernel learning.
\newblock {\em JMLR - Proceedings Track}, 5:336--343, 2009.

\bibitem{Lichtenauer20082040}
J.F. Lichtenauer, E.A. Hendriks, and M.J.T. Reinders.
\newblock Sign language recognition by combining statistical dtw and
  independent classification.
\newblock {\em IEEE Transactions on Pattern Analysis and Machine Intelligence},
  30(11):2040--2046, 2008.

\bibitem{DBLP:journals/pami/LingJ07}
Haibin Ling and David~W. Jacobs.
\newblock Shape classification using the inner-distance.
\newblock {\em {IEEE} Trans. Pattern Anal. Mach. Intell.}, 29(2):286--299,
  2007.

\bibitem{Liwicki20121624}
S.~Liwicki, S.~Zafeiriou, G.~Tzimiropoulos, and M.~Pantic.
\newblock Efficient online subspace learning with an indefinite kernel for
  visual tracking and recognition.
\newblock {\em IEEE Transactions on Neural Networks and Learning Systems},
  23(10):1624--1636, 2012.

\bibitem{Lu30082005}
F.~Lu, S.~Keles abd S.~J.~Wright, and G.~Wahba.
\newblock Framework for kernel regularization with application to protein
  clustering.
\newblock {\em Proceedings of the National Academy of Sciences of the United
  States of America}, 102(35):12332--12337, 2005.

\bibitem{Schleif2010h}
E.~Mwebaze, P.~Schneider, F.-M. Schleif, J.R. Aduwo, J.A. Quinn, S.~Haase,
  T.~Villmann, and M.~Biehl.
\newblock Divergence based classification in learning vector quantization.
\newblock {\em NeuroComputing}, 74:1429--1435, 2010.

\bibitem{neuhaus}
M.~Neuhaus and H.~Bunke.
\newblock Edit distance based kernel functions for structural pattern
  classification.
\newblock {\em Pattern Recognition}, 39(10):1852--1863, 2006.

\bibitem{Nguyen2013691}
N.Q. Nguyen, C.K. Abbey, and M.F. Insana.
\newblock Objective assessment of sonographic: Quality ii acquisition
  information spectrum.
\newblock {\em IEEE Transactions on Medical Imaging}, 32(4):691--698, 2013.

\bibitem{ny_orig}
E.~J. Nystr\"om.
\newblock {\"U}ber die praktische {A}ufl\"osung von {I}ntegralgleichungen mit
  {A}nwendungen auf {R}andwertaufgaben.
\newblock {\em Acta Mathematica}, 54(1):185--204, 1930.

\bibitem{Ong2004639}
C.S. Ong, X.~Mary, S.~Canu, and A.J. Smola.
\newblock Learning with non-positive kernels.
\newblock pages 639--646, 2004.

\bibitem{Pekalska2005a}
E.~Pekalska and R.~Duin.
\newblock {\em The dissimilarity representation for pattern recognition}.
\newblock World Scientific, 2005.

\bibitem{Haasdonk2009a}
Elsbieta Pekalska and Bernard Haasdonk.
\newblock Kernel discriminant analysis for positive definite and indefinite
  kernels.
\newblock {\em IEEE Transactions on Pattern Analysis and Machine Intelligence},
  31(6):1017--1032, 2009.

\bibitem{DBLP:journals/tsmc/PekalskaD08}
Elzbieta Pekalska and Robert P.~W. Duin.
\newblock Beyond traditional kernels: Classification in two dissimilarity-based
  representation spaces.
\newblock {\em IEEE Transactions on Systems, Man, and Cybernetics, Part C},
  38(6):729--744, 2008.

\bibitem{DBLP:conf/sspr/PekalskaDGB04}
Elzbieta Pekalska, Robert P.~W. Duin, Simon G{\"u}nter, and Horst Bunke.
\newblock On not making dissimilarities euclidean.
\newblock In {\em Structural, Syntactic, and Statistical Pattern Recognition,
  Joint IAPR International Workshops, SSPR 2004 and SPR 2004, Lisbon, Portugal,
  August 18-20, 2004 Proceedings}, pages 1145--1154, 2004.

\bibitem{DBLP:journals/pr/PekalskaDP06}
Elzbieta Pekalska, Robert P.~W. Duin, and Pavel Pacl{\'{\i}}k.
\newblock Prototype selection for dissimilarity-based classifiers.
\newblock {\em Pattern Recognition}, 39(2):189--208, 2006.

\bibitem{DBLP:journals/jmlr/PekalskaPD01}
Elzbieta Pekalska, Pavel Pacl\'{\i}k, and Robert P.~W. Duin.
\newblock A generalized kernel approach to dissimilarity-based classification.
\newblock {\em Journal of Machine Learning Research}, 2:175--211, 2001.

\bibitem{Platt:2005}
J.~Platt.
\newblock Fastmap, metricmap, and landmark mds are all nystr\"om algorithms,
  2005.

\bibitem{Schleif2014e}
F.-M. Schleif.
\newblock Proximity learning for non-standard big data.
\newblock In {\em Proceedings of ESANN 2014}, pages 359--364, 2014.

\bibitem{Schleif2013b}
F.-M. Schleif and A.~Gisbrecht.
\newblock Data analysis of (non-)metric proximities at linear costs.
\newblock In {\em Proceedings of SIMBAD 2013}, pages 59--74, 2013.

\bibitem{Cristianini2004a}
J.~Shawe-Taylor and N.~Cristianini.
\newblock {\em Kernel Methods for Pattern Analysis and Discovery}.
\newblock Cambridge University Press, 2004.

\bibitem{DBLP:conf/icml/SiHD14}
Si~Si, Cho{-}Jui Hsieh, and Inderjit~S. Dhillon.
\newblock Memory efficient kernel approximation.
\newblock In {\em Proceedings of the 31th International Conference on Machine
  Learning, {ICML} 2014, Beijing, China, 21-26 June 2014}, volume~32 of {\em
  {JMLR} Proceedings}, pages 701--709. JMLR.org, 2014.

\bibitem{citeulike:668527}
T.~F. Smith and M.~S. Waterman.
\newblock {Identification of common molecular subsequences.}
\newblock {\em Journal of molecular biology}, 147(1):195--197, March 1981.

\bibitem{JMLR:v14:talwalkar13a}
Ameet Talwalkar, Sanjiv Kumar, Mehryar Mohri, and Henry Rowley.
\newblock Large-scale svd and manifold learning.
\newblock {\em Journal of Machine Learning Research}, 14:3129--3152, 2013.

\bibitem{DBLP:conf/icml/TsangKK07}
Ivor~W. Tsang, Andr{\'a}s Kocsor, and James~T. Kwok.
\newblock Simpler core vector machines with enclosing balls.
\newblock In {\em Machine Learning, Proceedings of the Twenty-Fourth
  International Conference (ICML 2007), Corvallis, Oregon, USA, June 20-24,
  2007}, pages 911--918, 2007.

\bibitem{vanderMeer20063}
F.~v.~d. Meer.
\newblock The effectiveness of spectral similarity measures for the analysis of
  hyperspectral imagery.
\newblock {\em International Journal of Applied Earth Observation and
  Geoinformation}, 8(1):3--17, 2006.

\bibitem{vapnik2000nature}
V.N. Vapnik.
\newblock {\em The nature of statistical learning theory}.
\newblock Statistics for engineering and information science. Springer, 2000.

\bibitem{DBLP:conf/colt/LuxburgBB04}
Ulrike von Luxburg, Olivier Bousquet, and Mikhail Belkin.
\newblock On the convergence of spectral clustering on random samples: The
  normalized case.
\newblock In {\em Learning Theory, 17th Annual Conference on Learning Theory,
  COLT 2004, Banff, Canada, July 1-4, 2004, Proceedings}, pages 457--471, 2004.

\bibitem{DBLP:conf/kdd/WangZQZ14}
Shusen Wang, Chao Zhang, Hui Qian, and Zhihua Zhang.
\newblock Improving the modified nystr{\"{o}}m method using spectral shifting.
\newblock In Sofus~A. Macskassy, Claudia Perlich, Jure Leskovec, Wei Wang, and
  Rayid Ghani, editors, {\em The 20th {ACM} {SIGKDD} International Conference
  on Knowledge Discovery and Data Mining, {KDD} '14, New York, NY, {USA} -
  August 24 - 27, 2014}, pages 611--620. {ACM}, 2014.

\bibitem{wang2013improving}
Shusen Wang and Zhihua Zhang.
\newblock Improving {CUR} matrix decomposition and the {N}ystr\"om
  approximation via adaptive sampling.
\newblock {\em Journal of Machine Learning Research}, 14:2729--2769, 2013.

\bibitem{DBLP:conf/aistats/WangZ14}
Shusen Wang and Zhihua Zhang.
\newblock Efficient algorithms and error analysis for the modified nystrom
  method.
\newblock In {\em Proceedings of the Seventeenth International Conference on
  Artificial Intelligence and Statistics, {AISTATS} 2014, Reykjavik, Iceland,
  April 22-25, 2014}, volume~33 of {\em {JMLR} Proceedings}, pages 996--1004.
  JMLR.org, 2014.

\bibitem{werner}
Dirk Werner.
\newblock {\em Funktionalanalysis}.
\newblock Springer-Verlag Berlin Heidelberg, 2011.

\bibitem{DBLP:conf/nips/WilliamsS00}
Christopher K.~I. Williams and Matthias Seeger.
\newblock Using the nystr{\"o}m method to speed up kernel machines.
\newblock In {\em Advances in Neural Information Processing Systems 13, Papers
  from Neural Information Processing Systems (NIPS) 2000, Denver, CO, USA},
  pages 682--688, 2000.

\bibitem{DBLP:journals/tnn/ZhangK10a}
Kai Zhang and James~T. Kwok.
\newblock Clustered {N}ystr{\"o}m method for large scale manifold learning and
  dimension reduction.
\newblock {\em IEEE Transactions on Neural Networks}, 21(10):1576--1587, 2010.

\bibitem{DBLP:journals/jmlr/ZhangLWM12}
Kai Zhang, Liang Lan, Zhuang Wang, and Fabian Moerchen.
\newblock Scaling up kernel svm on limited resources: A low-rank linearization
  approach.
\newblock {\em JMLR - Proceedings Track}, 22:1425--1434, 2012.

\bibitem{DBLP:conf/icml/ZhangTK08}
Kai Zhang, Ivor~W. Tsang, and James~T. Kwok.
\newblock Improved {N}ystr{\"o}m low-rank approximation and error analysis.
\newblock In William~W. Cohen, Andrew McCallum, and Sam~T. Roweis, editors,
  {\em ICML}, volume 307 of {\em ACM International Conference Proceeding
  Series}, pages 1232--1239. ACM, 2008.

\end{thebibliography}
\end{document}